\documentclass[aps,prd,showpacs,showkeys,amsmath,amssymb]{revtex4} 
\usepackage[english]{babel}
\usepackage{graphicx,epsfig,float,subfigure,amssymb,amsmath,amsthm}

\newcommand{\beq}{\begin{equation}}
\newcommand{\eeq}{\end{equation}}
\newcommand{\bea}{\begin{eqnarray}}
\newcommand{\eea}{\end{eqnarray}}
\newcommand{\meio}{{}^1\!/{}_{\!2}}
\newcommand{\tmeio}{{}^3\!/{}_{\!2}}
\newcommand{\tquarto}{{}^3\!/{}_{\!4}}
\newcommand{\quarto}{{}^1\!/{}_{\!4}}

\DeclareMathOperator{\sech}{sech}%
\DeclareMathOperator{\cosech}{cosech}

\begin{document}

\title{Solutions to position-dependent mass quantum mechanics for a new class of hyperbolic potentials}

\author{H. R. Christiansen${}^{*,}{}^\dagger$   and M. S. Cunha${}^*$}
\affiliation{${}^*$Grupo de F\'{\i}sica Te\'orica, State University of Ceara (UECE), Av. Paranjana 1700, 60740-903 Fortaleza - CE, Brazil\\
${}^\dagger$ \!\!\!\! State University Vale do Acara\'{u}, Av. da Universidade 850, 62040-370 Sobral - CE, Brazil}

\begin{abstract}
We analytically solve the position-dependent mass (\textit{PDM}) 1D Schr\"odinger equation for a new class of hyperbolic potentials $V_q^p(x) = -V_0\frac{\sinh^px}{\cosh^qx}, \, p= -2, 0, \dots q$ [see C. A. Downing, J. Math. Phys. 54 072101 (2013)] among which several hyperbolic single- and double-wells. For a solitonic mass distribution, $m(x)=m_0\,\sech^2(x)$, we obtain exact analytic solutions to the resulting differential equations. For several members of the class, the quantum mechanical problems map into confluent Heun differential equations. The \textit{PDM} Poschl-Teller potential is considered and exactly solved as a particular case.
\end{abstract}

\email[]{hugo.christiansen@uece.br, marcony.cunha@uece.br}
\maketitle

\section{Introduction\label{intro}}

The dynamics of quantum  systems has been extensively studied
in connection with countless problems of physics, from low to high energies,
since the beginnings of the last century.
Non-relativistic quantum particles are accurately described by
a Schr\"odinger equation whose difficulties exclusively depend on the potential
function associated with the external background.
Mathematically, the problem consists in dealing
with the resulting differential equation which, for stationary systems,
winds up in an eigenvalue problem in a Hilbert space \cite{cohen}.
Exact solutions of solvable potentials have been generally found in terms
of polynomials, exponential, trigonometric, hyperbolic, hypergeometric and even elliptic
functions \cite{morse,eckart,rosenmorse,manningrosen,poschlteller,manning,ma47,
bargmann,scarf,ff61,cimento62,cimento64,CPM66,natanzon79,nieto78,prl83,gino84}.

Although the number of exactly solvable potentials has been slowly growing up along the last
 nighty years, as soon as we switch the traditionally constant mass into a position-dependent
distribution the mathematical challenge becomes dramatically different.
Notwithstanding, a few phenomenological potentials with appropriate mass distributions have been
worked out in recent years and their solutions have also been found proportional to special functions
\cite{ardasever2011,iran,severtezcan,midyaroy2009,bagchi}.

Historically, a crucial motivation for the PDM approximation came from many-body problems in solid
state and condensed matter physics \cite{wannier,slater,luttingerkohn,BDD,bastard81,vonroos,einehem88,levy95}.
One example is the effective-mass model related to the envelope-function approximation used
to study the dynamics of electrons in semiconductor heterostructures \cite{BDD,zhukro83,weisvin91}.

In fact, particles with spatially dependent distributions of mass have been investigated
in several relevant subjects of low-energy physics related to the understanding of the electronic
properties of semiconductor heterostructures, crystal-growth
techniques \cite{gorawill,bastardetal,bastard88},
quantum wells and quantum dots \cite{dots},
Helium clusters \cite{helium}, graded crystals \cite{prl93},
quantum liquids \cite{qliq}, etc., and has recently been shown to be appropriate for the description
of growth-intended geometrical nanowire structures suffering size variations, impurities, dislocations,
and geometry imperfections \cite{willatzenlassen2007}.

In connection with the mathematical aspects of the PDM issue,  here we focus on the class
of solutions of the effective PDM Schr\"odinger equation for a family of potentials
\begin{equation}
V_{p,q}(x) = -V_0\frac{\sinh^p(x/d)}{\cosh^q(x/d)}, \, p= -2, 0, \dots q
\end{equation}
reported a few months ago  in the usual constant mass framework \cite{downing} (see Sec. \ref{sec:potential} for details).
There, it is claimed that for this potentials class the equation cannot
be necessarily transformed into any hypergeometric special case \cite{abramowitz} as traditionally covet,
and it is shown that for  $q=4, 6$ the Schr\"odinger equation is equivalent to a \textit{confluent Heun equation} \cite{heun},
more difficult to handle \cite{maier,maier192,conflu,ronveaux,hille,slavyanov,hounkonnou,fiziev}
and not so well-known as the above mentioned.
Actually, the author manages to compute the low lying spectrum for the case $(p,q)=(4,6)$. 

In the present work we analyze this new family of potentials with a noteworthy modification, namely
a nonuniform mass for the quantum particle, and we further explore several members of this class.
This phenomenological upgrade
 of course induces highly nontrivial consequences in the associated differential equations and
yields new physical models. Our goal is to
analytically treat these resulting equations and find their solutions.
We do it in several representative $(p, q)$ cases
and find again confluent Heun equations (with new due entries)
even where there were just hypergeometric ones for a constant mass.
After a number of simple transformations we were able to obtain exact expression for their eigenstates.

In the next section, \ref{sec:ordering}, we digress about the question of which
suitable differential operator one should consider
to conveniently modify the Schr\"odinger equation for PDM. In Sec. \ref{sec:potential} we obtain an effective potential
in a convenient space and in Sec. \ref{sec:family}  we consider the $V_{(p,q)}(x)$
family of potentials and discuss one by one several $(p, q)$ representatives among which the hyperbolic
\textit{PDM} Poschl-Teller wells and the hyperbolic \textit{PDM} double-well.
Explicit wave-functions and eigenvalues are presented along with some selected graphics.
In all the cases we take the constant mass limit and compare it with the results
of Ref. \cite{downing}. The final remarks are drawn in Sec.\ref{sec:conclusion}.

\section{The PDM Schr\"odinger equation \label{sec:generalpdm}}
In this section we discuss the operational aspect of the effective hamiltonian of the quantum system
when the mass of the particle is assumed to depend on the spatial coordinate.

\subsection{Digression on the momentum operator\label{sec:ordering}}

Since we want to consider a position-dependent mass $m(x)$ in the Schr\"odinger equation,
an ordering problem arises from scratch with the momentum operator.
Clearly, the $x$-representation of the momentum,
$\hat P=-i\hbar\; d/dx$, demands considering derivatives not just on the wave-function
but also on the mass. The consequences of such inherent ordering dilemma are in fact serious
since a bunch of possibilities spreads out  \cite{bastard81,vonroos,einehem88,levy95,BDD,zhukro83,gorawill,shewell,morrowbron84,vonroofmavro,
morrow,galgeo,likhun,thomsenetc,young,einevol1,einevol2,tdlee,DA}

Instead of the ordinary kinetic-energy operator $\hat T_0= \frac 12m^{-1}\, \hat P^2$,
a fully comprehensive operator allowing all such ordering possibilities reads
\beq
  \hat T =\, \frac 14 \, \hat T_0\,+\frac 18\left\{\,\,\hat P^2\,m^{-1}(x)
  +\,  m^{\alpha }(x)\,\hat{P}\ m^{\beta}(x)\,\hat{P}\ m^{\gamma }(x)\,\,+\,m^{\gamma }(x)\,\hat{P}
\ m^{\beta }(x)\,\hat{P}\ m^{\alpha }(x)\, \right\},\label{Top}
\eeq
with  $\alpha +\beta +\gamma =-1$  \cite{vonroos}.
Taking into account the commutation rules of Quantum Mechanics, $[\hat X,\hat P]=i\hbar$,
we get
\begin{equation}
\hat T\,=\,\frac{1}{2\,m}\,\hat{P}^{2}\,+\,\frac{i\hbar }{2}\frac{1}{m^{2}}\frac{dm}{dx}
\,\,\hat{P}\,\,+\,U_{\rm K}\left( x\right),
\end{equation}
where
\beq
U_{\rm K}\left( x\right) =\frac{-\hbar ^{2}}{
4m^{3}(x)}\left[ \left( \alpha +\gamma -1\right) \frac{m(x)}{2}\left(\frac{
d^{2}m}{dx^{2}}\right)+\left(1-\alpha \gamma -\alpha -\gamma
\right) \left( \frac{dm}{dx}\right) ^{2}\right]\label{kinpot}
\end{equation}
is an effective potential of \textit{kinematic} origin. Potential $U_{\rm K}$ is thus
a source of an uncomfortable ambiguity in the hamiltonian of the physical problem
since its spatial shape depends on the values of $\alpha, \beta, \gamma$.

Nevertheless, if we also constrain
\begin{equation}
\alpha \,+\,\gamma \,=\,1\,=\alpha \,\gamma \,+\,\alpha\,+\,\gamma  \label{cond}
\end{equation}
the kinetic operator $\hat T$ becomes unique. As an extra bonus one finds that the first-try
symmetrized-Weyl order \cite{tdlee} (viz. $\alpha \,=\,\gamma=\,0$) gets excluded
in favor of the resulting, and most consensual, Ben-Daniel--Duke ordering \cite{BDD} which
is the one solution to the non-ambiguity condition Eq. (\ref{cond})
(viz. $\alpha =0$ and $\gamma=1$, or  $\alpha=1 $ and $\gamma =0$).

In any case, although free of the uncertainties of the kinematic potential (\ref{kinpot}),
the new effective Schr\"odinger equation now includes a first order derivative term. For an arbitrary external
potential $V(x)$ the PDM-Schr\"odinger equation thus reads

\beq %
\left \{ \frac{d^2}{dx^2} -\left(\frac{1}{m(x)} \frac{d m(x)}{dx}\right) \frac{d}{dx} +
\frac{2}{\hbar^2}\,m(x)\, \big[E-V(x)\big] \right \}
\psi(x) =0. \label{pdm}
\eeq
\\

Noticeably, not only the last term has been strongly modified from the ordinary Schr\"odinger equation but
the differential operator turned out to be dramatically changed. This will have of course deep consequences
on the physical wave solution of the system.

\subsection{The potentials\label{sec:potential}}

The potential functions we are dealing with are
\begin{equation}
V_{p,q}(x) = -V_0\frac{\sinh^p(x/d)}{\cosh^q(x/d)}, \, p= -2, 0, \dots q
\end{equation}
where the depth $V_0$ and width $d$, and the family parameters $p$ and $q$,
determine a plethora of possible shapes.

Here we adopt the following smooth effective mass distribution
\beq m(x)=m_0\, \sech^2(x/d) \label{massa} \eeq
which mimics a solitonic mass density
(see e.g. \cite{iran} and \cite{bagchi})
familiar in effective models of condensed matter and low energy nuclear physics.
The scale parameter $d$ widens the shape of the effective mass as it gets larger. Inversely,  the mass and
energy scales decrease as $d$ grow up, see Eq. (\ref{eqSch}).
Now, the resulting Schr\"odinger equation reads
\beq
\psi'' (x)+2 \tanh(x)
\psi'(x)+ \frac{2m_0d^2}{\hbar^2}(E-V_{p,q}(x))\, \sech^2(x) \psi(x)=0, \label{eqSch}
\eeq
where we shifted $x/d \rightarrow x$.
Trying the ansatz solution
\beq
\psi(x)=\cosh^{\nu}\!(x)\,\varphi(x), \label{transf} \eeq
Eq. (\ref{eqSch}) becomes
\beq
\varphi''(x)+2 (\nu +1) \tanh(x)\varphi'(x)+\left[\nu(\nu+2) \tanh^2\!x
+ \left(\nu +\frac{2m_0}{a^2\hbar^2}(E-V_{p,q}(x)) \right)\,\sech^2(x) \right] \varphi(x)=0.
\label{eqtransf}
\eeq
%
%
%
Now, we map the domain $(-\infty,\infty) \rightarrow (-\frac{\pi}{2},\frac{\pi}{2})$ by means of
a change of variable $\textsf{sech}\,x = \cos\,z,  \label{transf2} $
getting
\beq
\varphi''(z)+(2\nu+1)\tan(z)\varphi'(z)+\left[
\nu+\nu(\nu+2)\tan^2(z)+\frac{2m_0d^2}{\hbar^2}\left(E-\tilde{V}_{p,q}(z)\right) \right]\varphi(z)=0.
\eeq
Choosing $\nu=-1/2$ we can remove the first derivative  and obtain
\beq
\left[-\frac{d^2}{dz^2}+ \mathcal{V}_{p,q}(z)\right]\varphi(z)=\mathcal{E}\varphi(z) \label{schrodinger-cons}
\eeq
where
\beq
\mathcal{V}_{p,q}(z)=\frac{1}{2} +\frac{3}{4}\tan^2\!z -\mathcal{V}_0\ {\tan^{p}\!z}\,{\cos^{q}\!z}, \label{mathcalV}
\eeq
 $\mathcal{E}=\frac{2m_0d^2}{\hbar^2}E$ and $\mathcal{V}_0=\frac{2m_0d^2 }{\hbar^2} V_0$.

In this way, we transformed the differential equation (\ref{pdm}) into a true ordinary Schr\"odinger equation, viz. Eq.(\ref{schrodinger-cons}), which
 governs the dynamics of a particle of constant mass $m_0$ moving in an effective
potential free of any kinematical contribution. 
Now the dynamics is restricted to within
$z =(-\frac{\pi}{2},\frac{\pi}{2})$ and $\varphi(z=\pm \frac{\pi}{2})=0$.
Eventually, we can easily transform everything back to
the original variable and wave-function to obtain the real space solution.

\section{Analysis of solutions: Heun functions\label{sec:family}}

In this section we inspect one by one different members of this family of potentials allowing
a position-dependent mass in the Schr\"odinger-like differential equation (\ref{pdm}).
As we will see, this class of problems has complicated solutions most of which are Heun functions of
the confluent type. A confluent Heun differential equation arises from a general Heun
equation under a procedure which consists in making diverge three of its six parameters while holding
them proportional in a precise way. In its confluent form two or more regular singularities merge to
give rise to an irregular singularity \cite{conflu}.
Notice that, as demonstrated in \cite{maier}, a local Heun solution with four regular singular points
can be reduced to a Gauss hypergeometric function with only three regular singular points in some cases.
This will be explicitly shown in the first of the following family cases. The other constituents of the
class have irreducible confluent Heun solutions as we will see next.

\subsection{The case $(p,q)=(0,0)$ \label{sec:null}}

For $p=0$ and $q=0$ the potential $V_{(0,0)}(x)=-V_0$, which amounts to a
$V(x)=0$ case without loss of generality. Thus, in a sense it belongs to all the family members.
Although null, this is anyway a nontrivial situation as a result of the position-dependent mass.
We have already studied this issue in \cite{cunha-christiansen2013} so we shall review it quickly.
It will be helpful to show the relevant steps of our derivation and to
exhibit the analytical procedure and its connection with the Heun equation as well.

We start directly with Eq. (\ref{schrodinger-cons}). Upon a second transformation of variables, $y = \cos z$,
and with an ansatz solution $\varphi(y)=y^{-1/2}h_1(y)$, we obtain
\beq
h_1''(y)+\left( \frac{-1}{y}+\frac{\meio}{y-1}+\frac{\meio}{y+1}\right)
h_1'(y)+\frac{-k^2 y}{y (y-1) (y+1)}h_1(y) = 0\,\,
\label{eqh}
\eeq
where $0< y < 1$ and $k^2=2m_0d^2E/\hbar^2$.  
This equation belongs to the second order fuchsian class \cite{hille} and
can be recognized as a special case of the Heun equation \cite{ronveaux}
\beq
H''(y)+\left( \frac{\gamma}{y}+\frac{\delta}{y-1}+\frac{\varepsilon}{y-d} \right) H'(y)+\frac{\alpha \beta y-q}{y(y-1)(y-d)} H(y) =0\,\, \label{heun} \eeq
where the fuchsian relation $\alpha+\beta+1=\gamma+\delta+\varepsilon$ holds.
%
%
%
We  look for solutions around the singularity $y=1$, which corresponds to $x=0$ in the original space.
For this the characteristic exponents are $0$ and $1-\delta$ and thus
the two relevant linearly independent local Heun solutions to Eq.\ref{heun} are
\bea %
&&H^{(1)} = H(1-d,-q+\alpha \beta ,\alpha,\beta,\delta,\gamma; 1-y) \nonumber \\
&& H^{(2)}=(1-y)^{1-\delta}\! H[1-d,-q+(\delta-1)\gamma d + (\alpha-\delta+1)(\beta-\delta+1),\beta-\delta+1,
\alpha-\delta+1, 2- \delta, \gamma; 1-y].
\eea

\subsubsection{Even solutions}
Looking at Eq.(\ref{eqh}) and Eq.(\ref{heun})
we identify $d=-1$, $q=0$, $\alpha=-\meio +\meio\sqrt{1+4k^2}$, $\beta=-\meio
-\meio\sqrt{1+4k^2}$, $\gamma=-1$,  $\delta=\meio$ and $\varepsilon=\meio$
which result  in
\bea
h_1^{(1)}(y) = H\left(2,-k^2,\frac{-1+\sqrt{1+4k^2}}{2},
\frac{-1-\sqrt{1+4k^2}}{2},\meio,-1; 1-y\right) ~~~~\label{eqh_par}\\
 h_1^{(2)}(y) = (1\!-\!y)^{1/2}\!H\!\left(\!2,-\frac{3+4k^2}{4},
-\frac{\sqrt{1+4k^2}}{2},\frac{\sqrt{1+4k^2}}{2},\tmeio,\!-1;
1\!-\!y\right)~~
\eea
corresponding to
\beq \psi^{(1,2)}(x) = h_1^{(1,2)}(\sech x) \eeq
in the original space. The relevant boundary conditions are easily imposed in $z$
since the potential diverges to $+\infty$ in $z=\pm \frac{\pi}{2}$. This implies $\varphi(z=\pm
\frac{\pi}{2})=0$. The first solution is convergent only for $k^2=n(n+1)$ with $n=1,3,...$, and
the second solution is not acceptable because it is not differentiable at $z=0$ (see below).
\subsubsection{Odd solutions}

In order to obtain the odd solutions we first define
$\varphi(z)=\sin z\, \phi(z)$,
and then again $y=\cos z$ and $\phi(y)=y^{-1/2}h_2(y)$ which leads to
\beq
h_2''(y)+\left( \frac{-1}{y}+\frac{\tmeio}{y-1}+\frac{\tmeio}{y+1}\right)
h_2'(y)+\frac{-k^2 y}{y (y-1) (y+1)}h_2(y) =0. \label{eqh2}
\eeq 
As before, $h_2(y)$ are local Heun functions around $y=1$ (namely $z=0$ and $x=0$))
and are given by
\bea
 h_2^{(1)}(y) = H\left(2,-k^2,\frac{1+\sqrt{1+4k^2}}{2},
\frac{1-\sqrt{1+4k^2}}{2},\frac{3}{2},-1; 1-y\right) ~~~~~\label{eqh_impar}\\
 h_2^{(2)}(y) = (1\!-\!y)^{-1/2} H\!\left(\!2,\quarto- k^2,
-\frac{\sqrt{1+4k^2}}{2}, \frac{\sqrt{1+4k^2}}{2}, \frac{1}{2},-1;
1-y \right) ~~~
\eea
where the identification of parameters is
$\alpha_2=\frac{1+\sqrt{1+4k^2}}{2}$, $\beta_2=\frac{1-\sqrt{1+4k^2}}{2}$,
$\gamma_2=-1$ and $\delta_2 = \varepsilon_2 = \frac{3}{2}$.
In $z$ and $x$ space these read
\bea
\varphi^{(1,2)}(z) = \sin z\,\sec^{\!1/2}\!z~ h_2^{(1,2)}(\cos z)\\
\psi^{(1,2)}(x)=\tanh x~ h_2^{(1,2)}\sech x)~~~~~
\eea%
The first solutions, $\varphi^{(1)}(z)$ and $\psi^{(1)}(x)$, converge only
for $k^2=n(n+1)$, with $n>0$ even. Again, $\varphi^{(2)}(z)$ ($\psi^{(2)}(x)$)
is not differentiable at the origin and we discard it.

\subsubsection{Hypergeometric solutions}

As we advanced at the beginning of this section, in some cases the Heun equation can be reduced to a
ordinary hypergeometric equation \cite{maier}.
This is actually the $(p,q)=(0,0)$  case, where in terms of the space variables $z$ and $x$
the physically acceptable solutions can be respectively written as
\bea
\varphi_{\rm phys}^{(1)}(z) = \sec^{\!1/2}\!z\ {}_2F_1\left(\frac{-1+\sqrt{1+4k^2}}{4},
\frac{-1-\sqrt{1+4k^2}}{4}, \frac{1}{2}~ ;\, \sin^2\!z \right) \label{fis_sim_z}\\
 \varphi_{\rm phys}^{(2)}(z) = \sin\!z\sec^{\!1/2}\!z\ {}_2F_1\left(\frac{1+\sqrt{1+4k^2}}{4}, \frac{1-\sqrt{1+4k^2}}{4},
\frac{3}{2} \,;\, \sin^2\!z \right) \label{fis_asim_z} \eea
and
\bea
\psi_{\rm phys}^{(1)}(x) = {}_2F_1\!\left(\frac{-1+\sqrt{1+4k^2}}{4},
\frac{-1-\sqrt{1+4k^2}}{4}, \frac{1}{2};\, \tanh^2\!x \right) \label{fis_sim_x}~~~~\\
\psi_{\rm phys}^{(2)}(x) = \tanh\!x\ {}_2F_1\!\left(\frac{1+\sqrt{1+4k^2}}{4},
\frac{1-\sqrt{1+4k^2}}{4}, \frac{3}{2} ; \tanh^2\!x \right)\,\, \label{fis_asim_x}. \eea

\subsubsection{Analytic spectrum}

We now need to analyze the border conditions. Looking at Eq. (\ref{fis_sim_z}), we calculate
$\lim_{(z \rightarrow \pm \frac{\pi}{2})}\ \sec\,z = \infty$ and
\beq \lim_{z \rightarrow \pm \frac{\pi}{2}} {}_2F_1\left(\frac{-1+\sqrt{1+4k^2}}{4}, \frac{-1-\sqrt{1+4k^2}}{4},
\frac{1}{2}~ ;\, \sin^2\!z \right)
= \frac{\sqrt{\pi}}{\Gamma \left(\frac{3}{4}-\frac{1}{4}
\sqrt{1+4k^2} \right) \Gamma \left(\frac{3}{4}+\frac{1}{4}
\sqrt{1+4k^2}\right)}. \eeq
Since we expect $\varphi_{\rm phys}^{(1)}(z)$  to vanish at the frontier,
for finite $k$  the vanishing condition results
\beq
\frac{3}{4}-\frac{1}{4} \sqrt{1+4k^2} = -p, ~~ p =0, 1, ... \eeq
namely
\beq k^2 = (2p+1)^2+(2p+1). \label{odd}\eeq

Equivalently, for the antisymmetric solutions, Eq.(\ref{fis_asim_z}), we certify that
\beq \lim_{z \rightarrow \pm
\frac{\pi}{2}} {}_2F_1\left(\frac{1+\sqrt{1+4k^2}}{4}, \frac{1-\sqrt{1+4k^2}}{4},
\frac{3}{2}~ ;\, \sin^2\!z \right) =\frac{2\sqrt{\pi}}{k^2\Gamma \left(\frac{1}{4}-\frac{1}{4}
\sqrt{1+4k^2} \right) \Gamma \left(\frac{1}{4}+\frac{1}{4}
\sqrt{1+4k^2}\right)}. \eeq
and so we need
 \beq \frac{1}{4}-\frac{1}{4} \sqrt{1+4k^2} = -q, ~~ q =0, 1, ... \eeq
which, after regrouping reads
\beq
k^2=2q (2q+1). \label{even}
\eeq

Thus, the full energy spectrum analytically reads
\beq %
E = \frac{\hbar}{2m_0d^2}~ n(n+1),
\eeq %
with $n$ odd (even) for even (odd) solutions, respectively  ($n>0$, since energy cannot be zero).

The final expressions for symmetric and antisymmetric solutions (in $z$ and $x$ spaces)
are thus
\bea
\varphi_{\rm phys}^{(1)}(z) = \sec^{\!1/2}\!z \ {}_2F_1\left(\frac{n}{2},
-\frac{n+1}{2}, \frac{1}{2}~ ;\, \sin^2\!z \right) \label{fis_sim_z_n}\\
\varphi_{\rm phys}^{(2)}(z) = \sin\!z\sec^{\!1/2}\!z\
{}_2F_1\left(\frac{n+1}{2}, -\frac{n}{2}, \frac{3}{2} \,;\, \sin^2\!z
\right) \label{fis_asim_z_n} \eea
and
\bea
\psi_{\rm phys}^{(1)}(x) = {}_2F_1\!\left(\frac{n}{2},
-\frac{n+1}{2}, \frac{1}{2}~ ;\, \tanh^2\!x  \right) \label{fis_sim_x_n}~~~~\\
\psi_{\rm phys}^{(2)}(x) = \tanh\!x \ {}_2F_1\!\left(\frac{n+1}{2},
-\frac{n}{2}, \frac{3}{2} \,;\, \tanh^2\!x \right).
\label{fis_asim_x_n} 
\eea

\subsection{ The case  $(p,q)= (-2,0)$,  $V_{p,q}(x)= -V_0 \cosech^2\!x$\label{sec:-20}}

\begin{figure}[ht]
 \center
{\includegraphics[width=7.cm,height=5.5cm]{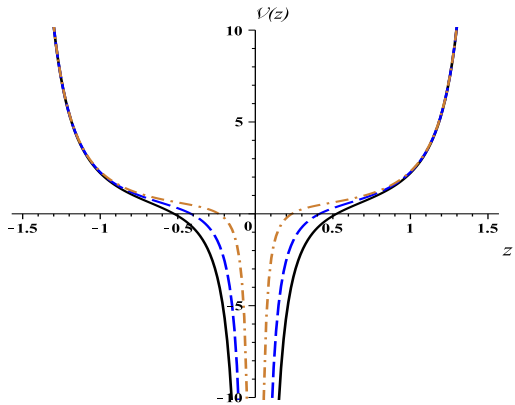}}
\caption{\label{fig_pot_z} The effective potential $\mathcal{V}_{(-2,0)}(z)$, Eq. (\ref{mathcalV}) for $\mathcal{V}_0=1/32$ (dot-dashed gold line),  $\mathcal{V}_0 = 1/8$ (dashed blue line), and $\mathcal{V}_0 =1/4$ (solid black line).}
 \end{figure}

\begin{figure}[ht]
 \center
{\includegraphics[width=7.cm,height=5.5cm]{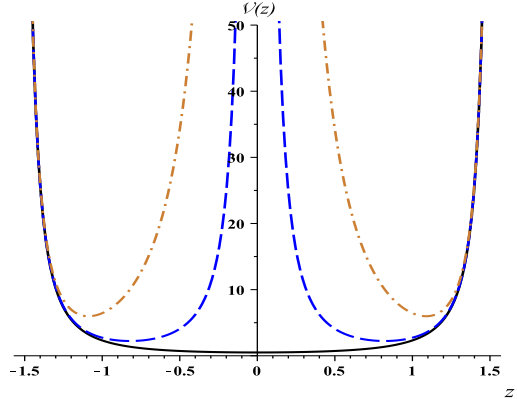}}
\caption{\label{fig_pot_z2} The effective potential $\mathcal{V}_{(-2,0)}(z)$, Eq. (\ref{mathcalV})  for $\mathcal{V}_0=-10$ (dot-dashed gold line),  $\mathcal{V}_0 = -1$ (dashed blue line), and $\mathcal{V}_0 =0$ (solid black line). Notice the double-well shape in z-space. The transition case $\mathcal{V}_0=0$ is the z-space single-well analyzed in Sec.\ref{sec:null}.}
 \end{figure}

This family member  $V_{(-2,0)}(x)= -V_0 \cosech^2\!x $ can be also shown to have solutions
which are factorable in hypergeometric functions.
Upon change of variables $\sech\,x = \cos\,z,$ the effective potential reads
\beq
\mathcal{V}_{(-2,0)}\!(z)   = \frac{1}{2}+\frac{3}{4}\tan^2(z)-\mathcal{V}_0\cot^2 \!z
\eeq
whose dependence on $\mathcal{V}_0 $ can be seen in Figs. \ref{fig_pot_z} and \ref{fig_pot_z2}.
It is interesting to note that this family member can represent very different potentials
in z space. For $\mathcal{V}_0>0$, potentials $\mathcal{V}_{(-2,0)}$ look like funnels while for  $\mathcal{V}_0<0$
potentials $\mathcal{V}_{(-2,0)}$ are infinite double-wells (the transition case $\mathcal{V}_0=0$ is the
infinite single-well analyzed in Sec.\ref{sec:null}).

For an ansatz solution of the form
\beq \varphi(z)=\sin^\mu\!(z) \cos^\nu\!(z)\, F(z)\eeq the z-space Schr\"odinger equation leads to
\beq
F''(z)+ 2\left(\mu \cot\!z-\nu \tan\!z\right) F'(z) +\left[(\mu^2-\mu+\mathcal{V}_0) \cot^2\!z -\mu-\nu-2\mu \nu +(\nu^2-\nu-\tquarto)\tan^2\!z+ \mathcal{E}-\meio \right]F(z) = 0,  \label{eqF}
\eeq
which, by choosing
\bea
\mu&=&\frac{1}{2} \pm \frac{1}{2}\sqrt{1-4\mathcal{V}_0} \label{eqmu}\\
\nu&=& \pm \frac{3}{2}, \label{eqnu}
\eea
can be  written as
\beq
F''(z)+ 2\left(\mu \cot\!z-\nu \tan\!z\right) F'(z) +\left( \mathcal{E}-\meio-\mu-\nu-2\mu \nu \right)F(z) = 0, \label{eqF2}
\eeq

Now,  further changing variables the definition of $\xi=\sin^2\!z$ yields
\beq
\xi(1-\xi)F''(\xi)+\left(\mu+\meio-(1+\mu+\nu)\,\xi\right)\,F'(\xi)+ \quarto \left(\mathcal{E}-\meio-\mu-\nu-2\mu\nu\right)F(\xi)=0,
\eeq
which can be recognized as a Gauss hypergeometric equation. Its linearly independent solutions are the hypergeometric series
${}_2F_1(a,b;c;\xi)$ and  $\xi^{1-c}{}_2F_1(a-c+1,b-c+1;2-c;\xi)$ (for $c\neq1$) \cite{abramowitz}.  In our case,
 the parameters have the following  values
\bea
a&=&\frac{\mu}{2}+\frac{\nu}{2}-\frac{1}{2}\sqrt{\mathcal{E}+\quarto-\mathcal{V}_0}  \\
b&=&\frac{\mu}{2}+\frac{\nu}{2}+\frac{1}{2}\sqrt{\mathcal{E}+\quarto-\mathcal{V}_0} \\
c&=&\mu+\frac{1}{2}.
\eea
Solutions to the $(p,q)= (-2,0)$ Schr\"odinger Eq. (\ref{schrodinger-cons}) therefore read
\bea
\varphi^{(1)}(z)&=&\sin^\mu\!z\cos^\nu\!z \,{}_2F_1(a,b;c;\sin^2\!z) \label{sol1}\\
\varphi^{(2)}(z)&=&\sin^\mu\!z\cos^\nu\!z \, (\sin^2\!z)^{1-c}{}_2F_1(a-c+1,b-c+1;2-c;\sin^2\!z). \label{sol2}
\eea
In order to cut the hypergeometric series we must compel respectively $a=-n$ and $a-c+1=-n$ with $n\,\epsilon\, N$.
This is because ${}_2F_1(a,b;c;\xi)$ and ${}_2F_1(a-c+1,b-c+1;2-c;\xi)$ diverge for $\xi=1$
while boundary conditions impose $\varphi(z=\pm\pi/2)=0$ in both cases. Note also that due to $\cos^\nu\!z$,
for a negative $\nu$ the solutions also diverge at the boundary so we set $\nu=3/2$.
Now, using Eqs.(\ref{eqmu}) and (\ref{eqnu}),  we analytically
obtain the  Eqs. (\ref{sol1}) and (\ref{sol2}) spectra
\beq
\mathcal{E}_n = 4(n+1)\left((n+1)\pm \sqrt{\quarto- \mathcal{V}_0}\right), \ \ \ \ n=0,1, 2\dots \label{-20spectrum}
\eeq
which are real for $\mathcal{V}_0\leqslant\quarto$.
However, assuming $\mu_+=\frac{1}{2}+\frac{1}{2}\sqrt{1-4\mathcal{V}_0}$, when $0<\mathcal{V}_0\leqslant\quarto$
the second solution is not physically acceptable for its derivative is divergent at the origin, and
when $\mathcal{V}_0<0$ the function itself is divergent at $z=0$. As a result, we shall discard this solution.
Therefore, the physical spectrum corresponds to the $^+$ case in Eq. (\ref{-20spectrum}) and the energy
eigenvalues are all positive. Now, for $\mu_-=\frac{1}{2}- \frac{1}{2}\sqrt{1-4\mathcal{V}_0}$ the same issues take place
with the first solution and we must keep the second one. But then, the $\pm$ spectra are interchanged.
In fact, the second solution with  $\mu_-$ happens to coincide with the first solution with $\mu_+$
so we can consider just $\varphi^{(1)}$ with $\mu_+$.

\begin{figure}[h]
  \center
 {\includegraphics[width=7.cm,height=5.5cm]{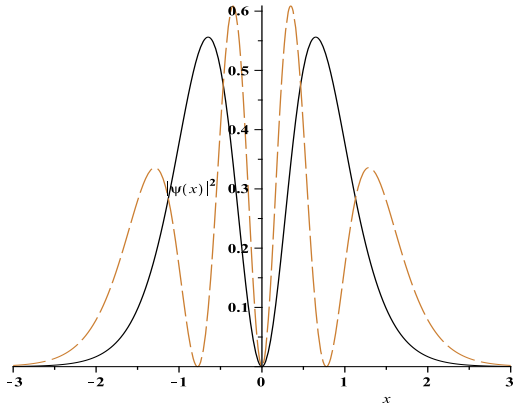}}
  \caption{\label{fig_n0_n1_v1_32_x} Plot of the normalized $\vert\psi_{\rm phys}(x)\vert^2$, Eq. (\ref{solx}), when $\mathcal{V}_0=1/32$
  for $n=0$ (solid) $\mathcal{E}_0=5.8708$, $n=1$ (dashed) $\mathcal{E}_1=19.7417$. See in Fig. \ref{fig_pot_z} the single-well effective potential $\mathcal{V}_{(-2,0)}(z)$.}
\end{figure}

\begin{figure}[h]
  \center
  {\includegraphics[width=7.cm,height=5.5cm]{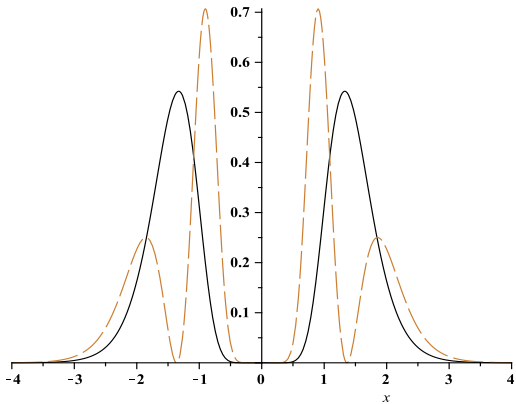}}
  \caption{\label{fig_n0_n1_v_-32_x} Plot of the normalized $\vert\psi_{\rm phys}(x)\vert^2$, Eq. (\ref{solx}), when $\mathcal{V}_0= -32$ for $n=0$ (solid) $\mathcal{E}_0=26.7156$, $n=1$ (dashed) $\mathcal{E}_1=61.4313$. See in Fig. \ref{fig_pot_z2} the double-well effective potential $\mathcal{V}_{(-2,0)}(z)$.}
\end{figure}

Thus, in $x$-space the solutions are eventually
\beq
\psi_{\rm phys}(x)= \tanh^{\mu_+}(x) \sech^{2}(x) {}_2F_1(a_+, b_+; c_+; \tanh^2(x)), \label{solx}
\eeq
whose probability densities are shown in Figs. \ref{fig_n0_n1_v1_32_x} and  \ref{fig_n0_n1_v_-32_x} for the low eigenvalues.
As above explained, neither the second expression, $\psi^{(2)}(x)=\tanh^{1-\mu}\!(x)\sech^{\nu+\frac{1}{2}} \!(x)\, {}_2F_1(a-c+1,b-c+1;2-c;\,\tanh^2\!(x))$, nor any of its Kummer's transformations \cite{maier192} are physically acceptable so we
shall not consider these solutions.

For $\mathcal{V}_0<0$ the effective potential is a double-well with a thin infinite barrier, see Fig. \ref{fig_pot_z2}.
The $z$-space solution $\varphi^{(1)}(z)$ exhibits a dumping shape about the origin, as expected for the physical solutions
of an ordinary Schr\"odinger equation. This same behavior is transferred to the $x$-space solution as shown in Fig. \ref{fig_n0_n1_v_-32_x}. The energy eigenvalues are higher than the $\mathcal{V}_0>0$ funnel-potentials ones, and
the higher they are the thinner is the dumping, as expected.

\subsection{The case  $(p,q)= (0,2)$, $V_{(p,q)}(x)= -V_0 \sech^2 x\,$ \label{sec:02}}

\begin{figure}[ht]
{\includegraphics[width=7.cm,height=5.5cm]{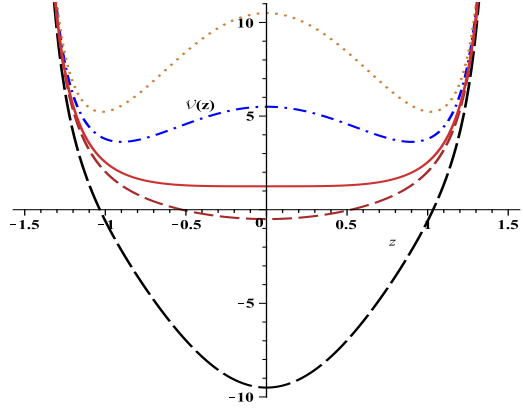}}
\caption{\label{figs_Vz_pq_02} Plot of $\mathcal{V}_{{(0,2)}}(z)$, Eq. (\ref{eqp0q2}), when
$\mathcal{V}_0=$ -10 (dotted line), -5 (dot-dashed), -3/4 (solid), 1 (dashed) and 10 (long-dashed).}
\end{figure}

This case is particularly appealing because the regular constant mass Schr\"odinger equation
is known as the Poschl-Teller equation \cite{poschlteller}. The regular Poschl-Teller potential $V_{{(0,2)}}(x)= -V_0 \sech^2 x$
looks like a soft finite single-well whose bound state solutions are just associate Legendre polynomials
$P_n^{\mu}(\tanh x), \, \mu=n, n-1, \dots 1$.
On the other hand, the effective z-space potential of the \textit{PDM} Poschl-Teller problem reads
\beq
\mathcal{V}_{{(0,2)}}(z) = \frac{1}{2}+\frac{3}{4}\tan^2\!z-\mathcal{V}_0\cos^2 \!z, \label{eqp0q2}
\eeq
and can be seen in Fig. \ref{figs_Vz_pq_02} for different values of $\mathcal{V}_0$.
Note the variation of the effective z-space potential from a single-well
to a double-well when $\mathcal{V}_0$ crosses the -3/4 value.

In order to solve the corresponding Schr\"odinger equation
\beq
-\frac{d^2 \varphi(z)}{dz^2}+\left [ \frac{1}{2} +\frac{3}{4}\tan^2\!z-\mathcal{V}_0\cos^2 \!z\right]\varphi(z) = \mathcal{E}\varphi(z), \label{schr_p0_q2} 
\eeq
we define $\varphi=\cos^\mu\!z H(z)$ so that we get an equation for $H(z)$

\beq
H''(z)-2\mu\tan(z) H'(z)+\left(\left(\mu^2-\mu -\tquarto\right) \tan^2\!z +\mathcal{E}
-\mu -\meio + \mathcal{V}_0\cos^2\!z \right]H(z) = 0.
\eeq
Upon a change of variables $\xi = \sin^2\!z$ and choosing $\mu = \tmeio$  it results
\beq
H''(\xi)+\left(\frac{\meio}{\xi}+ \frac{2}{\xi-1}\right) H'(\xi)+\frac{1}{\xi (\xi-1)} \left(-\frac{\mathcal{E}}{4}+\frac{1}{2}-\frac{\mathcal{V}_0}{4}+\frac{\mathcal{V}_0}{4}\xi \right) H(\xi) =0.
\eeq
Its solutions are therefore \textit{confluent Heun} functions
\bea
H^{(1)}(\xi) = Hc\!\left(0,-\frac{1}{2},1,\frac{\mathcal{V}_0}{4},\frac{1}{2}-\frac{\mathcal{E}+\mathcal{V}_0}{4};\xi \right)\\
H^{(2)}(\xi) = \xi^{\meio}Hc\!\left(0,\frac{1}{2},1,\frac{\mathcal{V}_0}{4},\frac{1}{2}-\frac{\mathcal{E}+\mathcal{V}_0}{4};\xi \right),
\eea
and the $z$ space solutions of Eq. (\ref{schrodinger-cons}) with potential (\ref{eqp0q2}) result
\bea
\varphi^{(1)}(z) = \cos^{\frac{3}{2}}\!z \,Hc\!\left(0,-\frac{1}{2},1,\frac{\mathcal{V}_0}{4},\frac{1}{2}-\frac{\mathcal{E}+\mathcal{V}_0}{4};\sin^2\!z \right)\,\\
\varphi^{(2)}(z) = \sin\!z \,\cos^{\frac{3}{2}}\!z\, Hc\!\left(0,\frac{1}{2},1,\frac{\mathcal{V}_0}{4},\frac{1}{2}-\frac{\mathcal{E}+\mathcal{V}_0}{4};\sin^2\!z \right).
\eea
Thus, in $x$ -space we have
\bea
\psi^{(1)}(x) = \sech^{2}\!x \,Hc\!\left(0,-\frac{1}{2},1,\frac{\mathcal{V}_0}{4},\frac{1}{2}-\frac{\mathcal{E}+\mathcal{V}_0}{4};\tanh^2\!x \right)
\label{psi1x pq02}\\
\psi^{(2)}(x) = \tanh\!x\, \sech^{2}\!x\, Hc\!\left(0,\frac{1}{2},1,\frac{\mathcal{V}_0}{4},\frac{1}{2}-\frac{\mathcal{E}+\mathcal{V}_0}{4};\tanh^2\!x \right).
\label{psi2x pq02}\eea



\begin{figure}[h]
\center
{\includegraphics[width=5.cm,height=4.5cm]{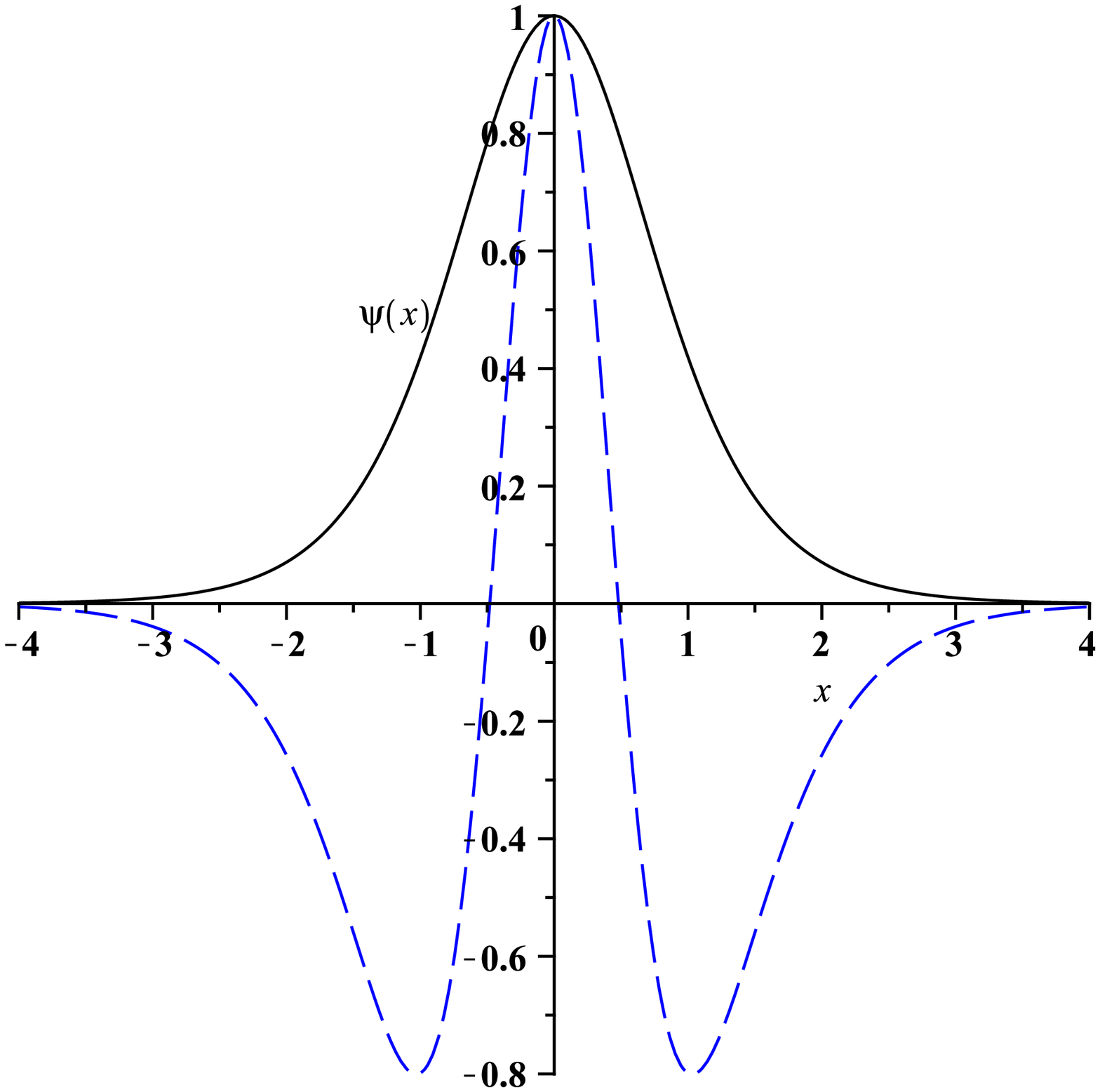}}
{\includegraphics[width=5.cm,height=4.5cm]{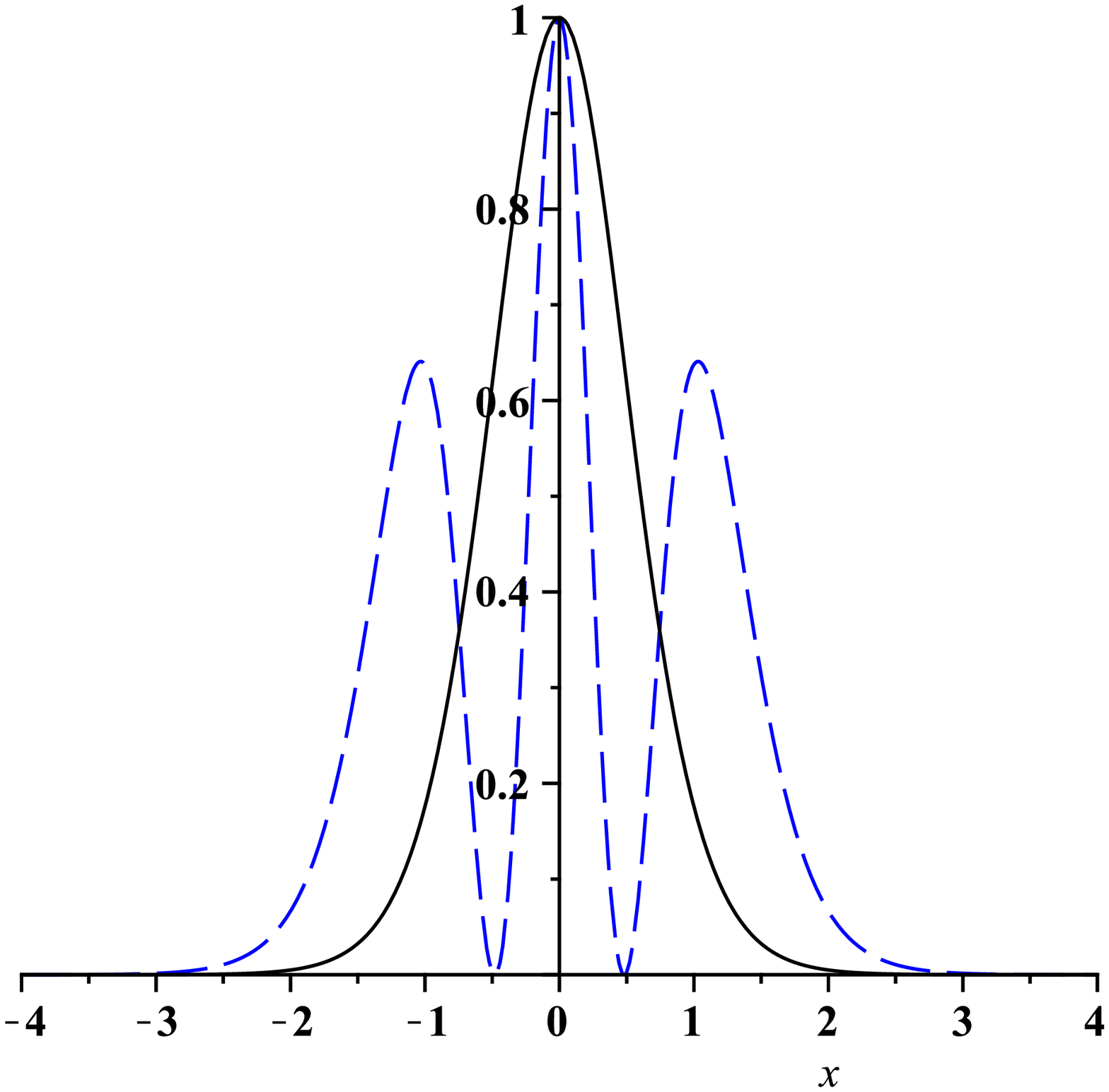}}
\center
{\includegraphics[width=5.cm,height=4.5cm]{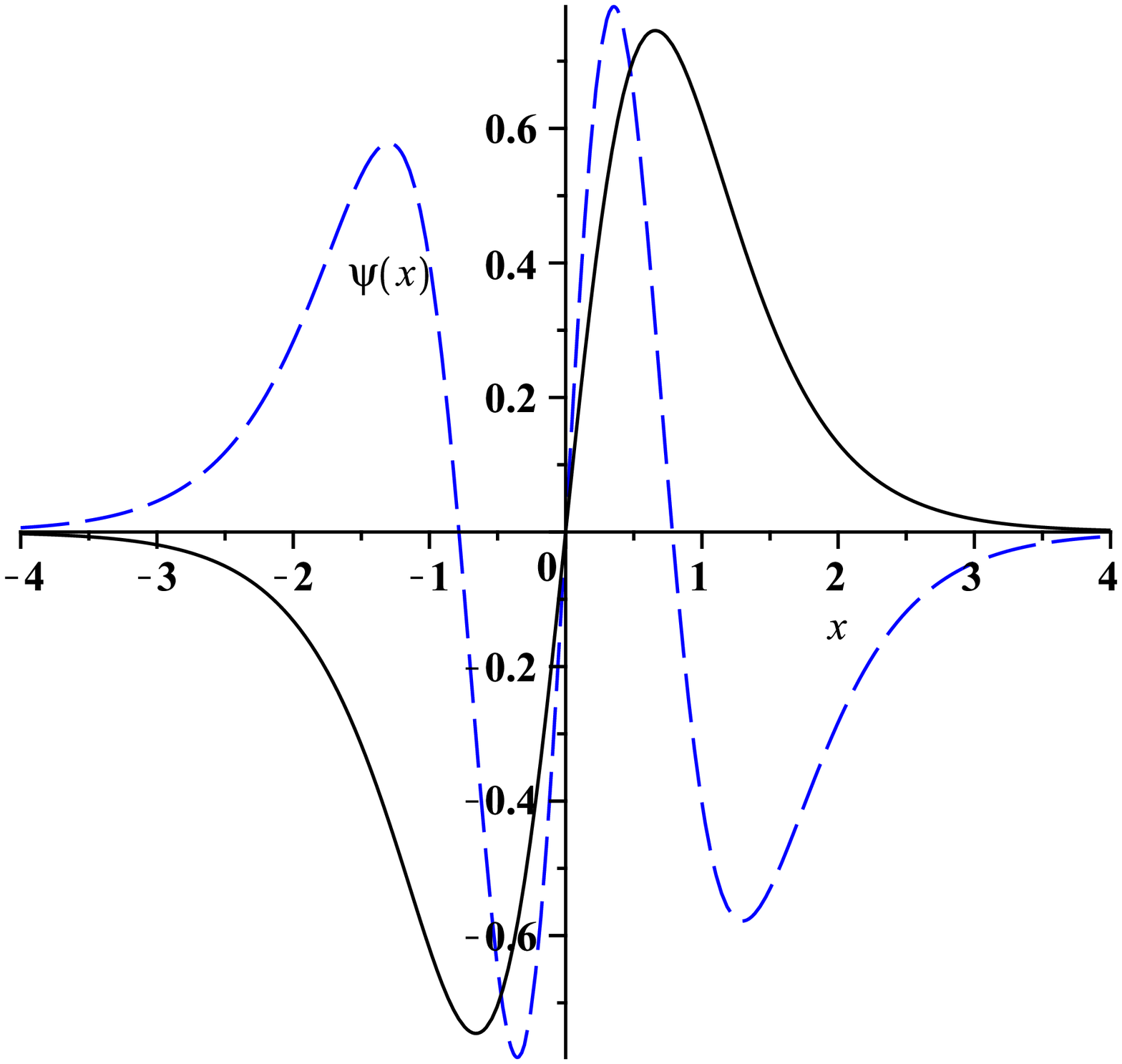}}
{\includegraphics[width=5.cm,height=4.5cm]{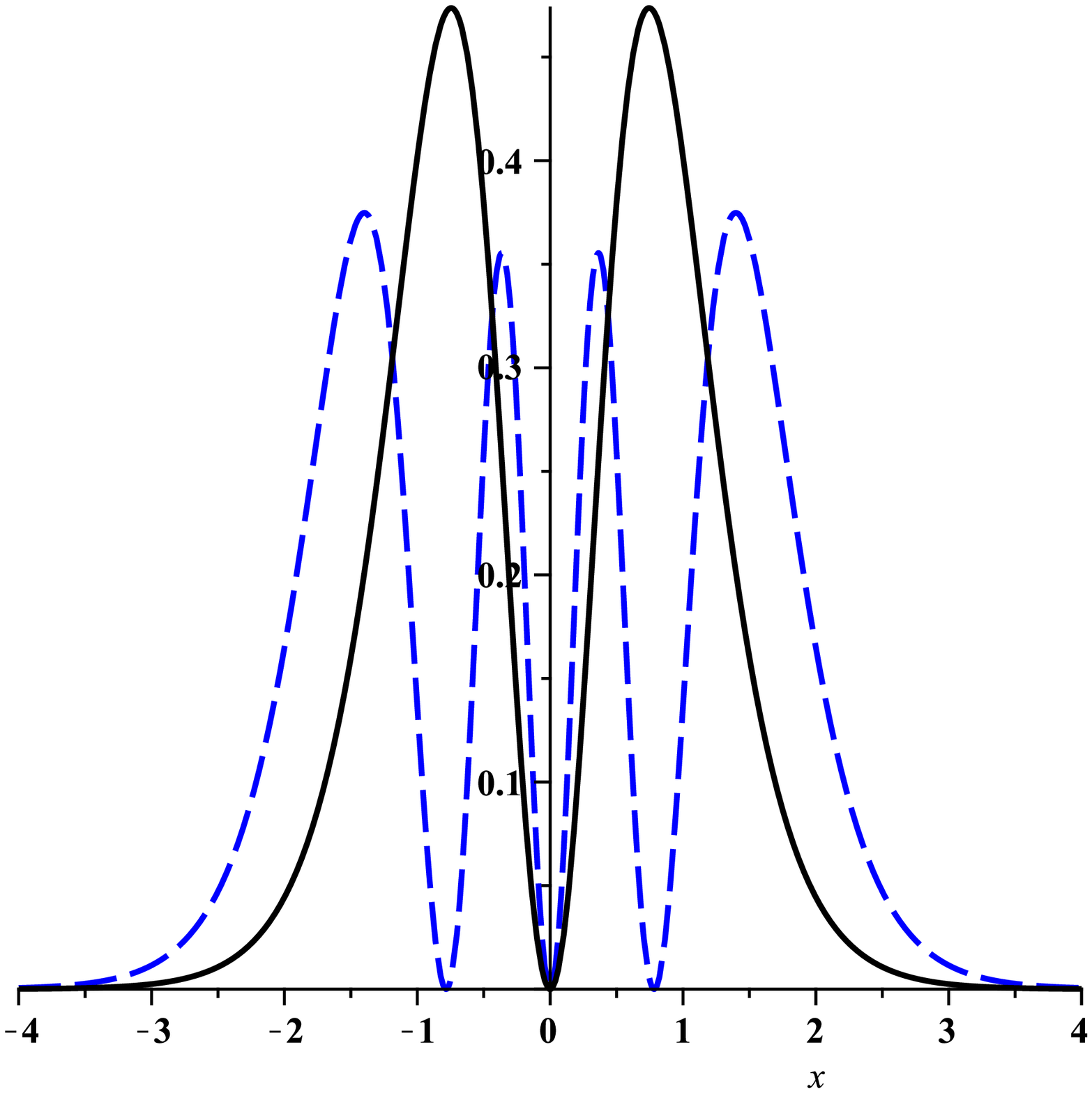}}
\caption{\label{pq_02_V50} Plot in x-space of normalized $\psi^{(1)}(x)$ (up left) and $\vert\psi^{(1)}(x)\vert^2$ (up right), Eq. (\ref{psi1x pq02}), for $\mathcal{E}_0 = 1.9749958440$ and $\mathcal{E}_2 = 11.983335512$, and $\psi^{(2)}(x)$ (down left) and $\vert\psi^{(2)}(x)\vert^2$ (down right), Eq. (\ref{psi2x pq02}),  for $\mathcal{E}_1$  = 5.982139325300; $\mathcal{E}_3$ = 19.9837682118. These are associated with the eigenfunctions $\varphi^{(1)}(z)$ and $\varphi^{(2)}(z)$ for a single-well potential $\mathcal{V}_{(0,2)}(z)$ with $\mathcal{V}_0=1/32$; this would correspond to a curve between the dashed and solid lines in Fig. \ref{figs_Vz_pq_02}.}
\end{figure}


\begin{figure}[h]
\center
{\includegraphics[width=5.cm,height=4.5cm]{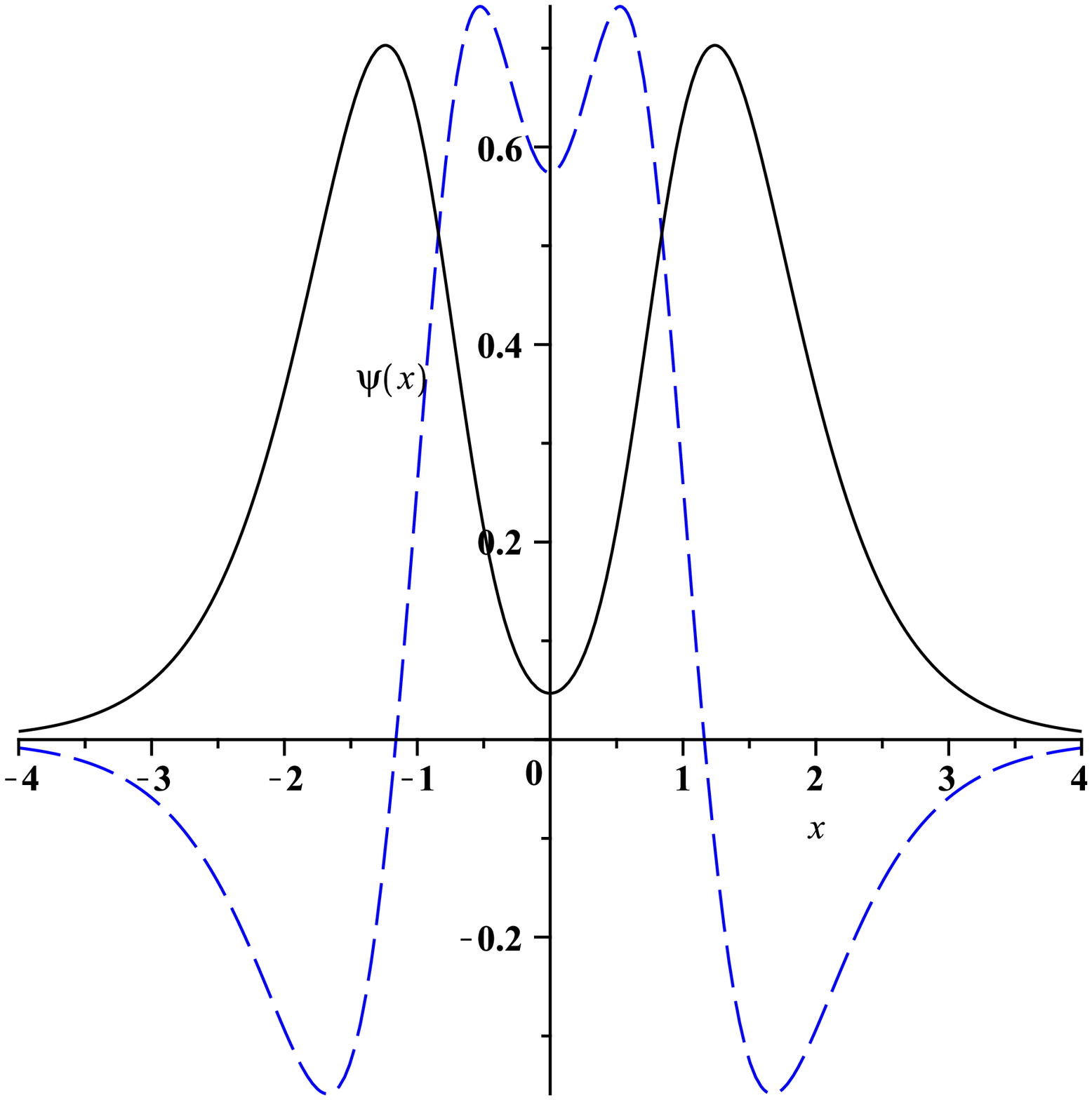}}
{\includegraphics[width=5.cm,height=4.5cm]{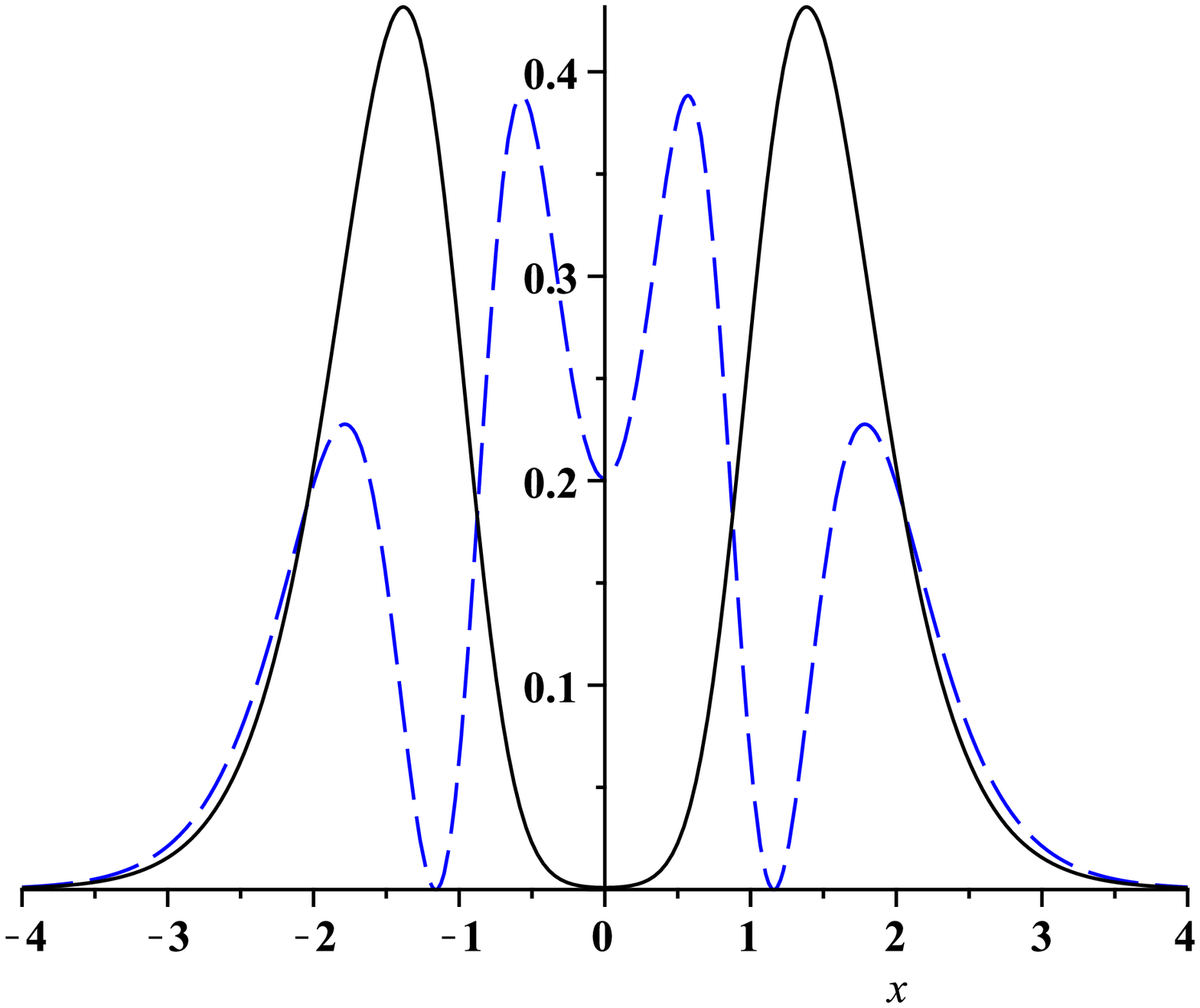}}
\center
{\includegraphics[width=5.cm,height=4.5cm]{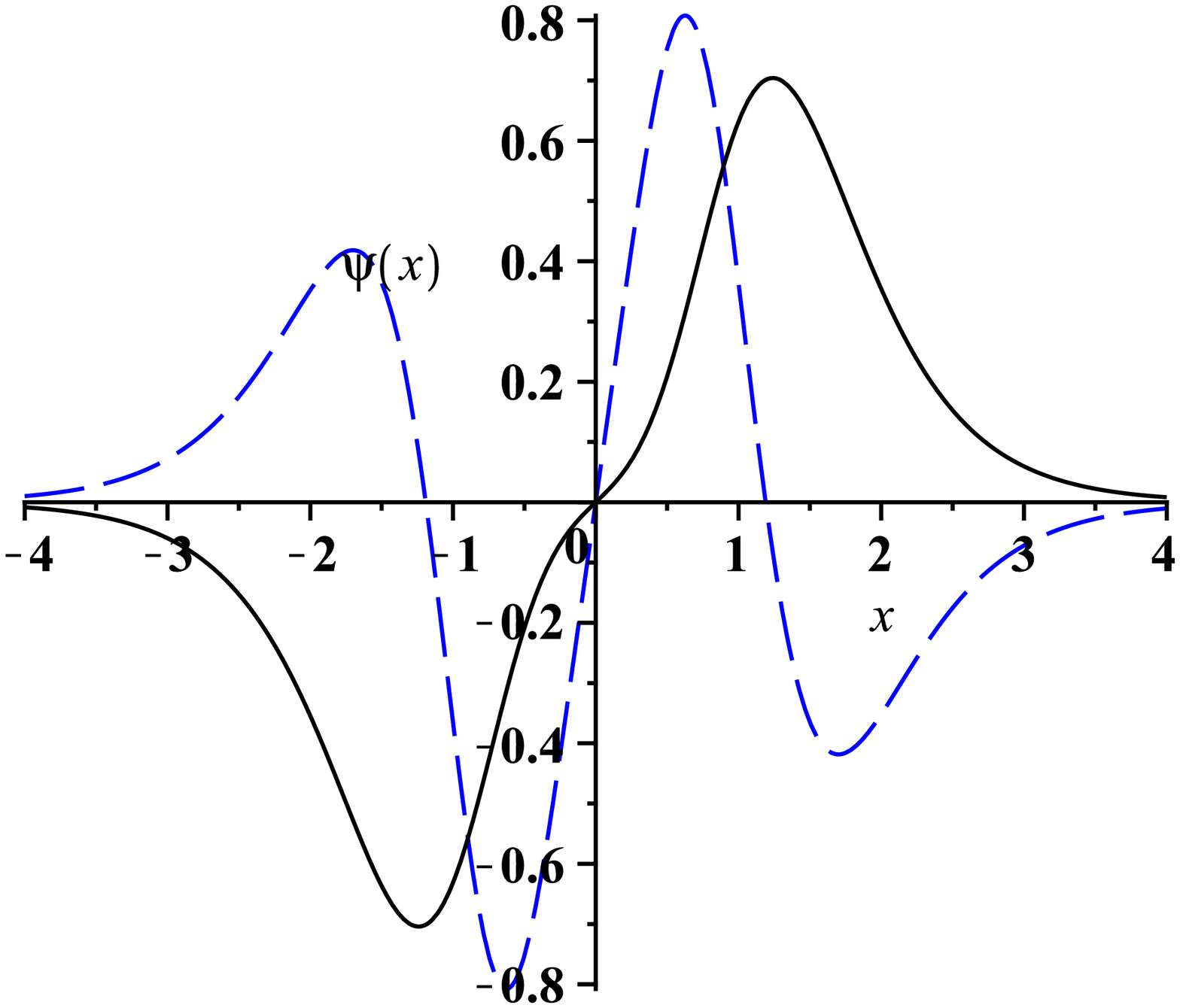}}
{\includegraphics[width=5.cm,height=4.5cm]{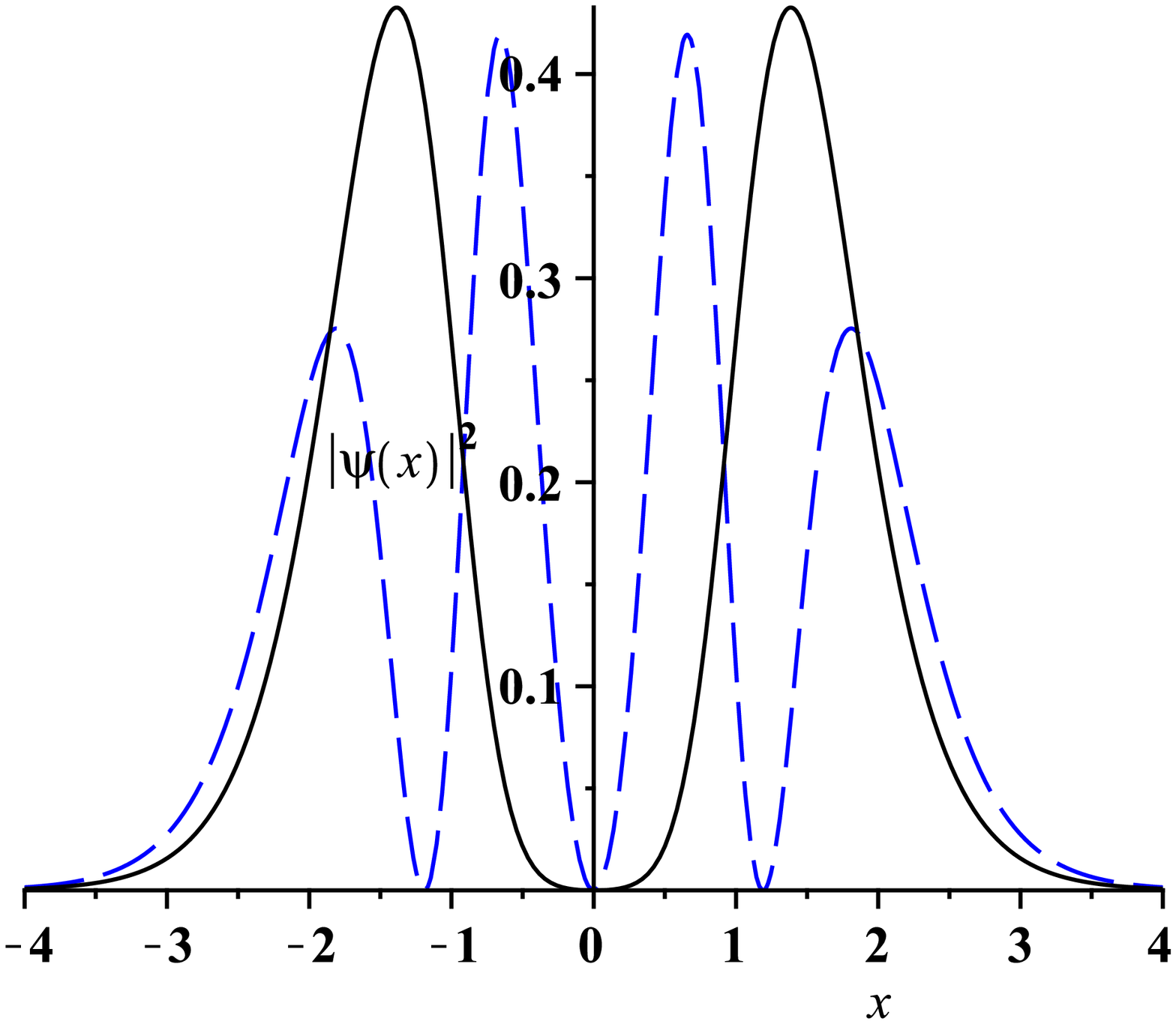}}
\caption{\label{pq_02_V-50} Plot in x-space of normalized $\psi^{(1)}(x)$ (up left) and $\vert\psi^{(1)}(x)\vert^2$ (up right), Eq. (\ref{psi1x pq02}), for $\mathcal{E}_0$ = 26.08773401704787 and $\mathcal{E}_2$ = 45.5205884270, and $\psi^{(2)}(x)$ (down left) and $\vert\psi^{(2)}(x)\vert^2$ (down right), Eq. (\ref{psi2x pq02}),  for $\mathcal{E}_1 $= 26.10903614483 and $\mathcal{E}_3 $= 47.56773146428. These are associated with the eigenfunctions $\varphi^{(1)}(z)$ and $\varphi^{(2)}(z)$ for a double-well potential $\mathcal{V}_{(0,2)}(z)$ with $\mathcal{V}_0=-50$; this would correspond to a curve above the dotted line in Fig. \ref{figs_Vz_pq_02}.}
\end{figure}

It is noteworthy the fact that the PDM effect is dramatic in every aspect.
For example, a potential barrier given by $V_{{(0,2)}}(x), \, V_0<0$, acts as an effective confining double-well
potential $\mathcal{V}_{(0,2)}(z)$  once the full differential equation is considered, see Fig. \ref{figs_Vz_pq_02}
 (dot and dot-dashed lines). The corresponding x-space wave-function behavior and probability distribution can be seen in the
associated eigenstates, Eqs. (\ref{psi1x pq02}) and (\ref{psi2x pq02}),  as shown in Fig. \ref{pq_02_V-50}.

\subsection{A special $(p,q)= (2,0)$ case,  $V(x)= -V_0\sinh^2 x\,$\label{sec:sech2}}

The inverse case, $(p,q)= (2,0)$,  $V(x)= -V_0\sinh^2 x\,$
does not exactly belong to the family defined in Sec. \ref{sec:potential} (since
here $p >q$) but it looks similar to its members and is also an interesting hyperbolic single-well.
So,  it is appropriate to briefly show its solutions here \cite{cunha-christiansen2013}.
Note that $V_{(2,0)}(z)=-V_0\tan^2 z$ which for the particular choice $V_0={3a^2\hbar^2}/{8m_0}$ makes
${\cal V}_{(2,0)}(z)$ trivial; see Eq. (\ref{mathcalV}).
The exact solutions to the corresponding PDM problem can be then easily obtained
\beq\varphi(z)=C_1\sqrt{\frac{2}{\pi}}\cos\big[(2n+1)z\big]+C_2\sqrt{\frac{2}{\pi}}\sin\left(2n z\right).\label{20z}
\eeq
For vanishing boundary conditions at $z=\pm \pi/2$, the energy is quantized as
\beq \mathcal{E}=\frac{a^2\hbar^2 \pi^2}{8m_0}n^2\eeq where $n\in \mathbb{N}$.
In $x$ space we obtain, from Eqs. (\ref{20z}) and (\ref{transf}),
\beq
\psi(x)=C_1 \sqrt{\frac{2}{\pi}}\sech^{1/2} \!x \,\sech\big[(2n+1) x\big] + C_2 \sqrt{\frac{2}{\pi}}\sech^{1/2} \!x \,\tanh(2n x).
\eeq


\subsection{The case $(p,q)= (2,4)$, $V_{(2,4)}(x)=-V_0 \sinh^2 x\sech^4 x\,$\label{sec:24}}

\begin{figure}[ht]
{\includegraphics[width=7.cm,height=5.5cm]{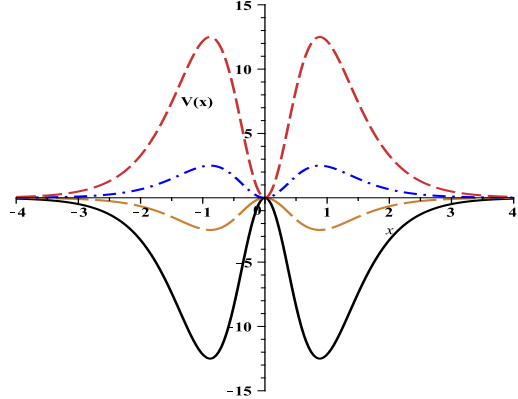}}
\caption{\label{pot_Vx_pq_24} Plot of ${V}_{{(2,4)}}(x)$ when ${V}_0=$ 50 (solid black line), 10 (dashed gold), -10 (dot-dashed blue), -50 (dashed red).}
\end{figure}

This potential, $V_{(2,4)}= -V_0 \sinh^2 x\sech^4 x\,$ (see Fig. \ref{pot_Vx_pq_24}), for ${V}_0>0$ represents a double-well in x-space regardless of the mass dependence. %

The z-space effective potential for the corresponding ordinary Schr\"odinger equation (\ref{schrodinger-cons}) results
\beq
\mathcal{V}_{{(0,2)}}(z) = \frac{1}{2}+\frac{3}{4}\tan^2\!z-\mathcal{V}_0 \sin^2\!z \cos^{2}\!z.
\label{Vz24}
\eeq
\begin{figure}[ht]
{\includegraphics[width=7.cm,height=5.5cm]{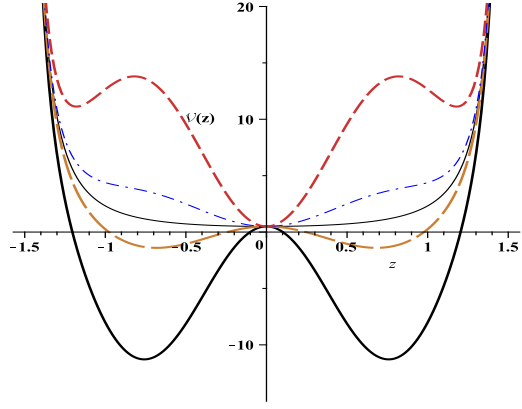}}
\caption{\label{pot_Vz_pq_24} Plot of $\mathcal{V}_{{(2,4)}}(z)$, Eq. (\ref{Vz24}), when  $\mathcal{V}_0=$ 50 (solid black line), 10 (dashed gold), -10 (dot-dashed blue), -50 (dashed red) associated with Fig. \ref{pot_Vx_pq_24}.}
\end{figure}
By defining $\xi=\sin^2\!z$, Eq. (\ref{schrodinger-cons}) becomes
\beq
\xi (\xi-1)\varphi''(\xi)+\left(\xi-\frac{1}{2} \right)\varphi'(\xi)-\frac{1}{4}\left(\mathcal{E}-\frac{1}{2}+\frac{3}{4}\frac{\xi}{\xi-1}-\mathcal{V}_0\xi (\xi-1)\right)\varphi(\xi)=0,
\eeq
which by means of $$\varphi(\xi)=(1-\xi)^\mu e^{\nu \xi} h(\xi)$$ gives the following differential equation for $h(\xi)$
\beq
h''(\xi)+\left(2\nu+\frac{2\mu+\meio}{\xi-1}+\frac{\meio}{\xi}\right) h'(\xi) + \frac{1}{\xi (\xi-1)} +
\left(-\frac{\mathcal{E}}{4}+\frac{1}{8}+ \frac{\mu-\nu}{2} + \nu(1+2\mu)\xi \right) h(\xi) = 0, \label{heunc_pq_24}
\eeq
provided
$$
\mu^2-\frac{\mu}{2}-\frac{3}{16}=0, \ \ \ \ \nu^2 + \frac{\mathcal{V}_0}{4}=0.
$$
The general solution to Eq. (\ref{heunc_pq_24}) is given by the following confluent Heun functions
\beq
h(\xi)=A_1\, Hc\!\left( 2\nu,\,-\frac{1}{2},\, 2\mu-\frac{1}{2},\,0,\,\frac{1}{2}-\frac{\mathcal{E}}{4};\,\xi \right)
+ A_2\,\xi^{\meio} Hc\!\left( 2\nu,\,\frac{1}{2},\, 2\mu-\frac{1}{2},\,0,\,\frac{1}{2}-\frac{\mathcal{E}}{4};\,\xi \right).
\eeq
The L.I. physical solutions are found for $\mu=\frac{3}{4}$ and $\nu=-\frac{\sqrt{-\mathcal{V}_0}}{2}$, resulting in
\bea
\varphi^{(1)}(z)= \,\cos^{\frac{3}{2}}\!z e^{-\frac{1}{2}\sqrt{-\mathcal{V}_0}\sin^ 2\!z} Hc\!\left( -\sqrt{-\mathcal{V}_0},\,-\frac{1}{2},\, 1,\,0,\,\frac{1}{2}-\frac{\mathcal{E}}{4};\,\sin^2\!z \right) \\
\varphi^{(2)}(z)=  \,\cos^{\frac{3}{2}}\!z\,e^{-\frac{1}{2}\sqrt{-\mathcal{V}_0}\sin^ 2\!z}\,\sin \!z\, Hc\!\left( -\sqrt{-\mathcal{V}_0},\,\frac{1}{2},\, 1,\,0,\,\frac{1}{2}-\frac{\mathcal{E}}{4};\,\sin^2\!z \right),
\eea
which in  $x$-space read (see Figs. \ref{pq_24_V50X} and \ref{pq_24_V-50X})
\bea
\psi^{(1)}(x)=A_1\,\sech^{2}\!x \,e^{-\frac{1}{2}\sqrt{-\mathcal{V}_0}\tanh^2\!x} Hc\!\left( -\sqrt{-\mathcal{V}_0},\,-\frac{1}{2},\, 1,\,0,\,\frac{1}{2}-\frac{\mathcal{E}}{4};\,\tanh^2\!x \right) \label{psisim_pq24}\\
\psi^{(2)}(x)= A_2\,\sech^{2}\!x\,e^{-\frac{1}{2}\sqrt{-\mathcal{V}_0}\tanh^ 2\!x}\,\tanh \!x\, Hc\!\left( -\sqrt{-\mathcal{V}_0},\,\frac{1}{2},\, 1,\,0,\,\frac{1}{2}-\frac{\mathcal{E}}{4};\,\tanh^2\!x \right).\label{psiantisim_pq24}
\eea

\begin{figure}[h]
\center
{\includegraphics[width=5.cm,height=4.5cm]{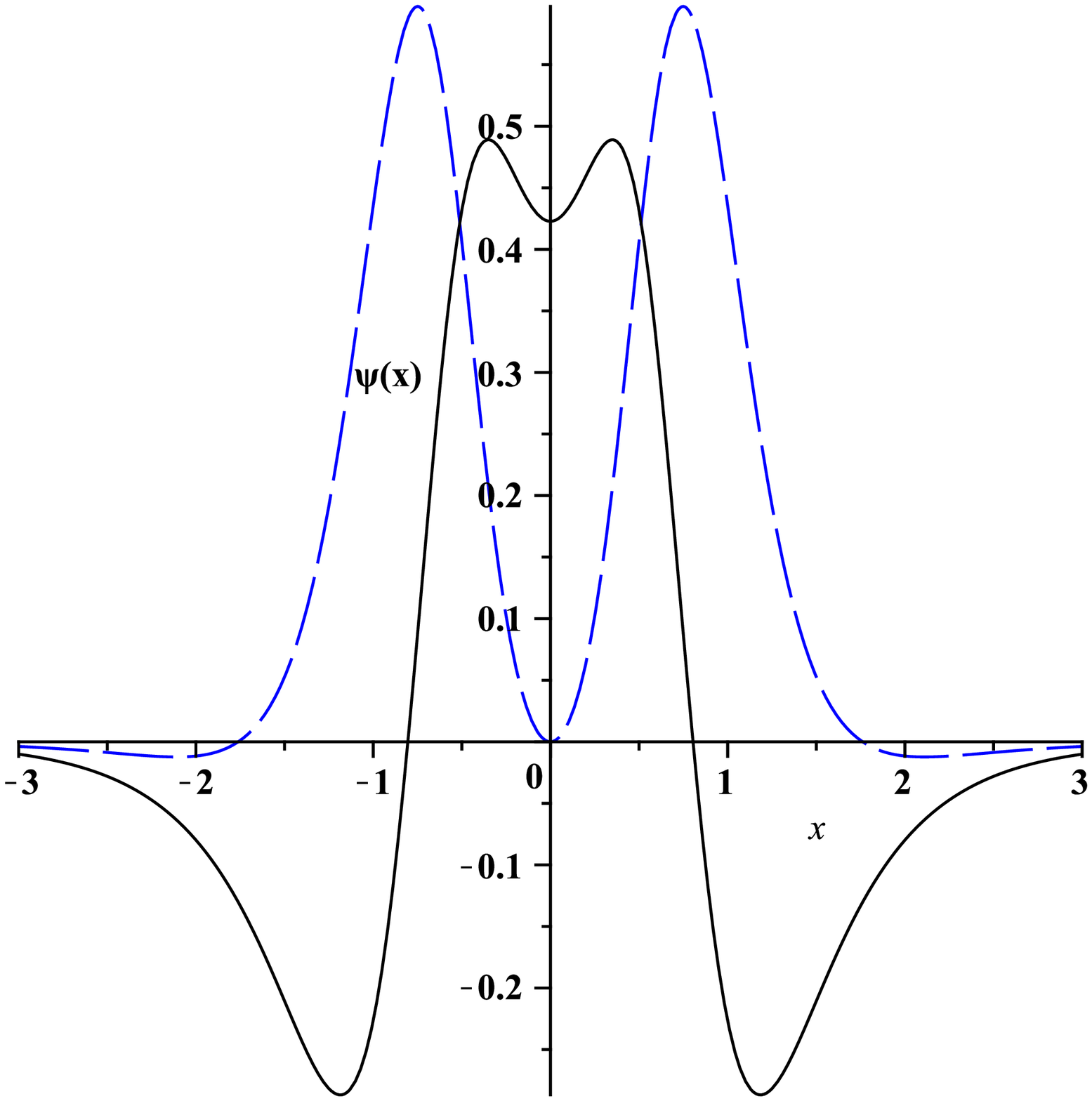}}
{\includegraphics[width=5.cm,height=4.5cm]{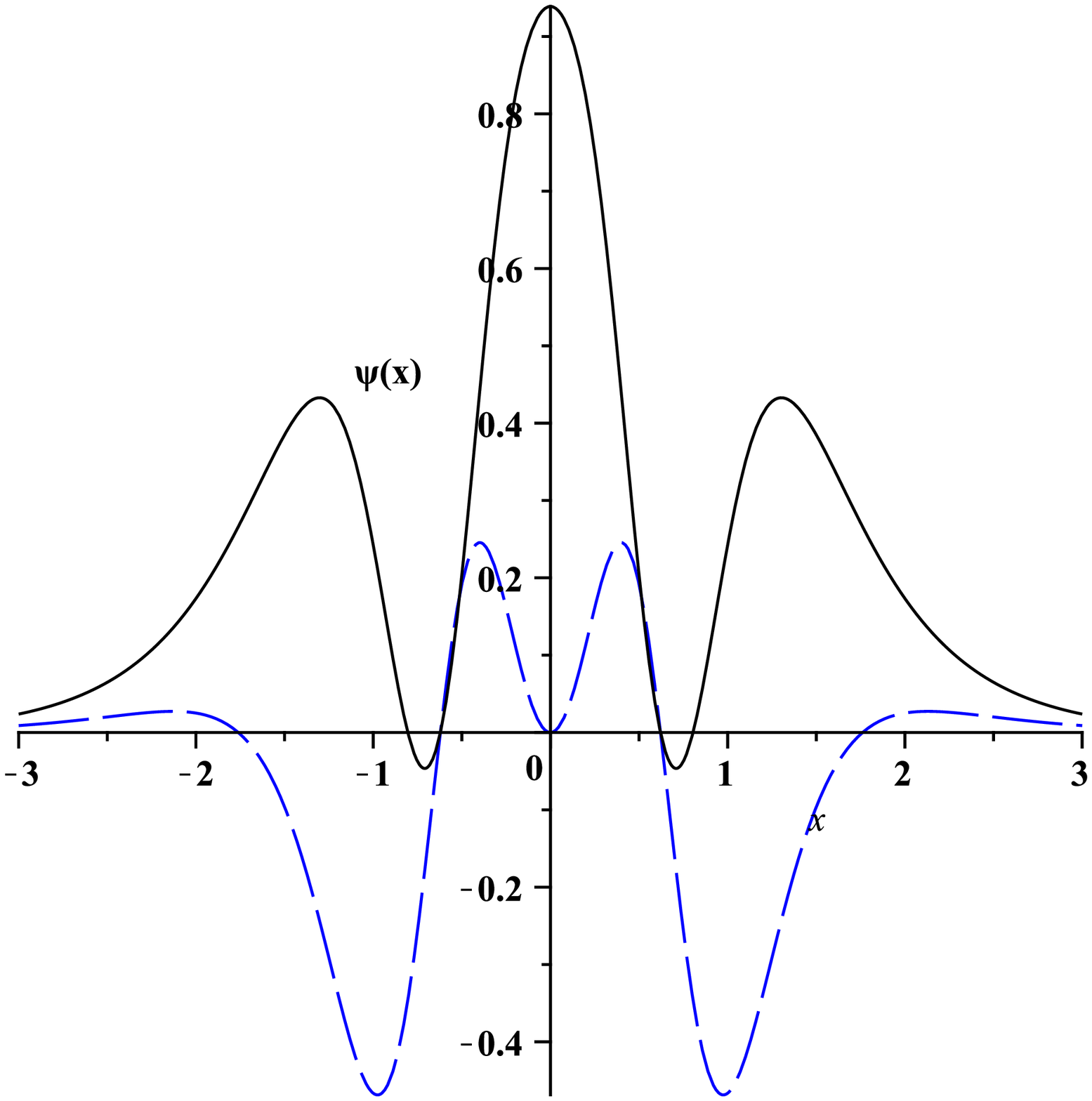}}
{\includegraphics[width=5.cm,height=4.5cm]{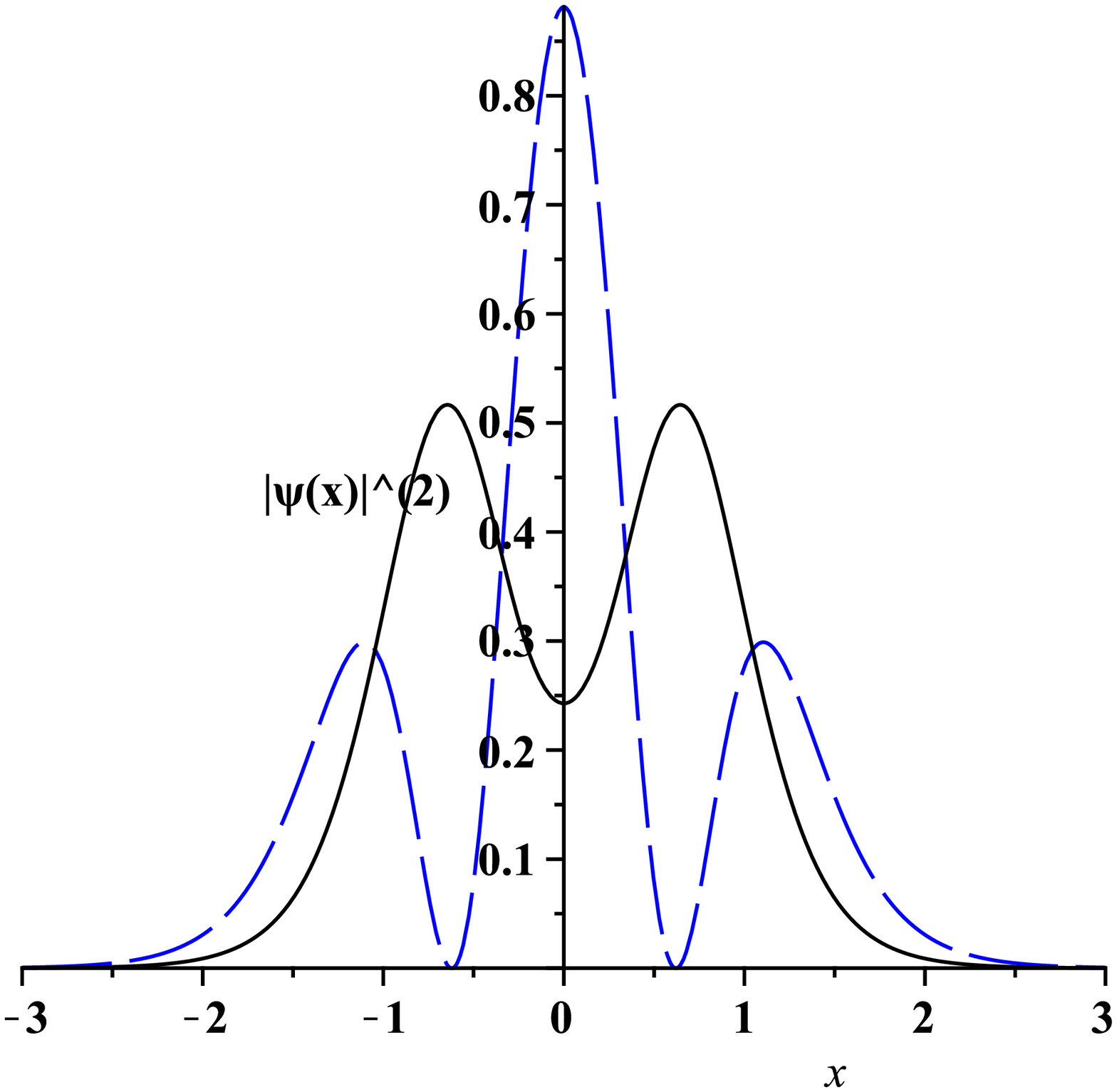}}
\center
{\includegraphics[width=5.cm,height=4.5cm]{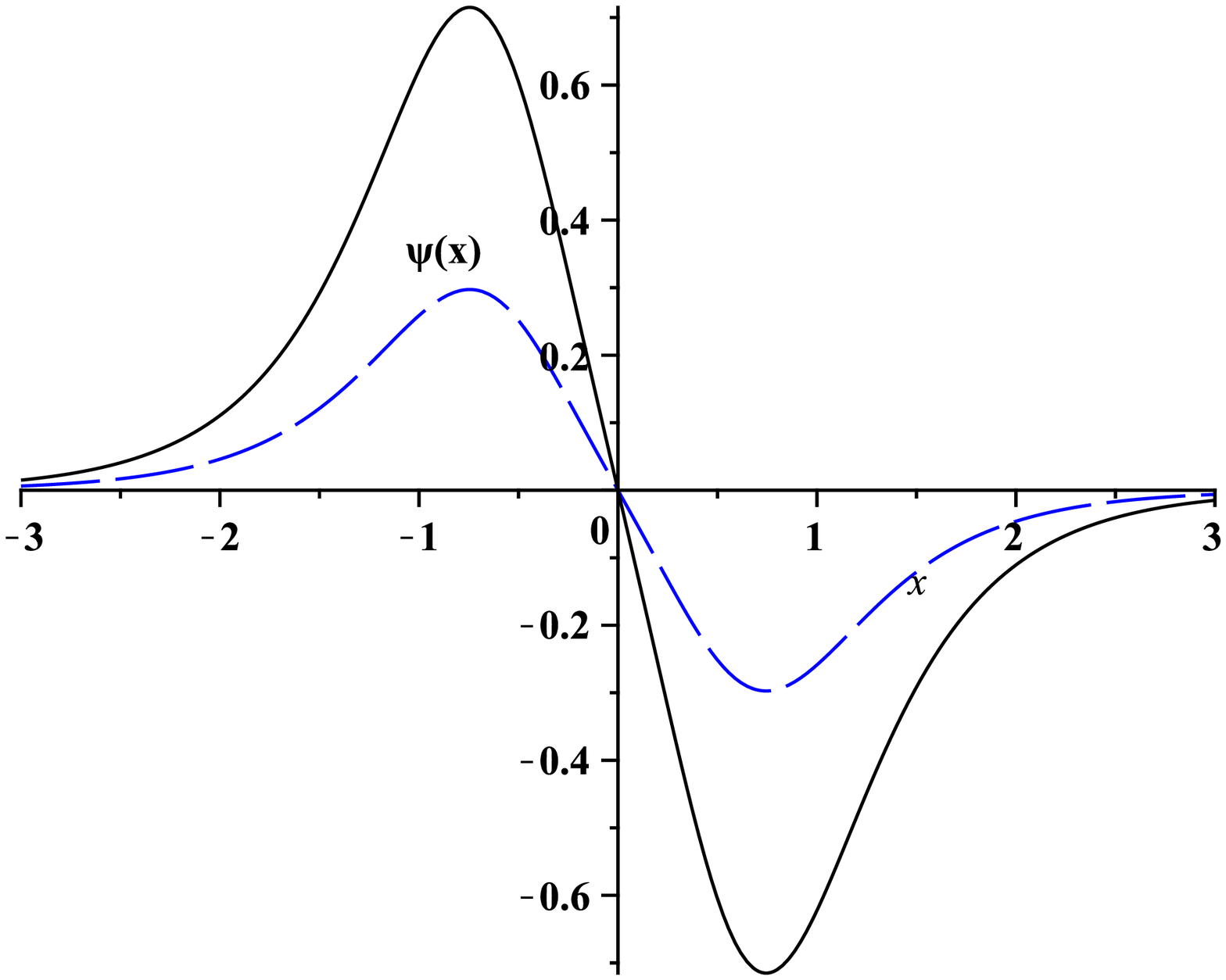}}
{\includegraphics[width=5.cm,height=4.5cm]{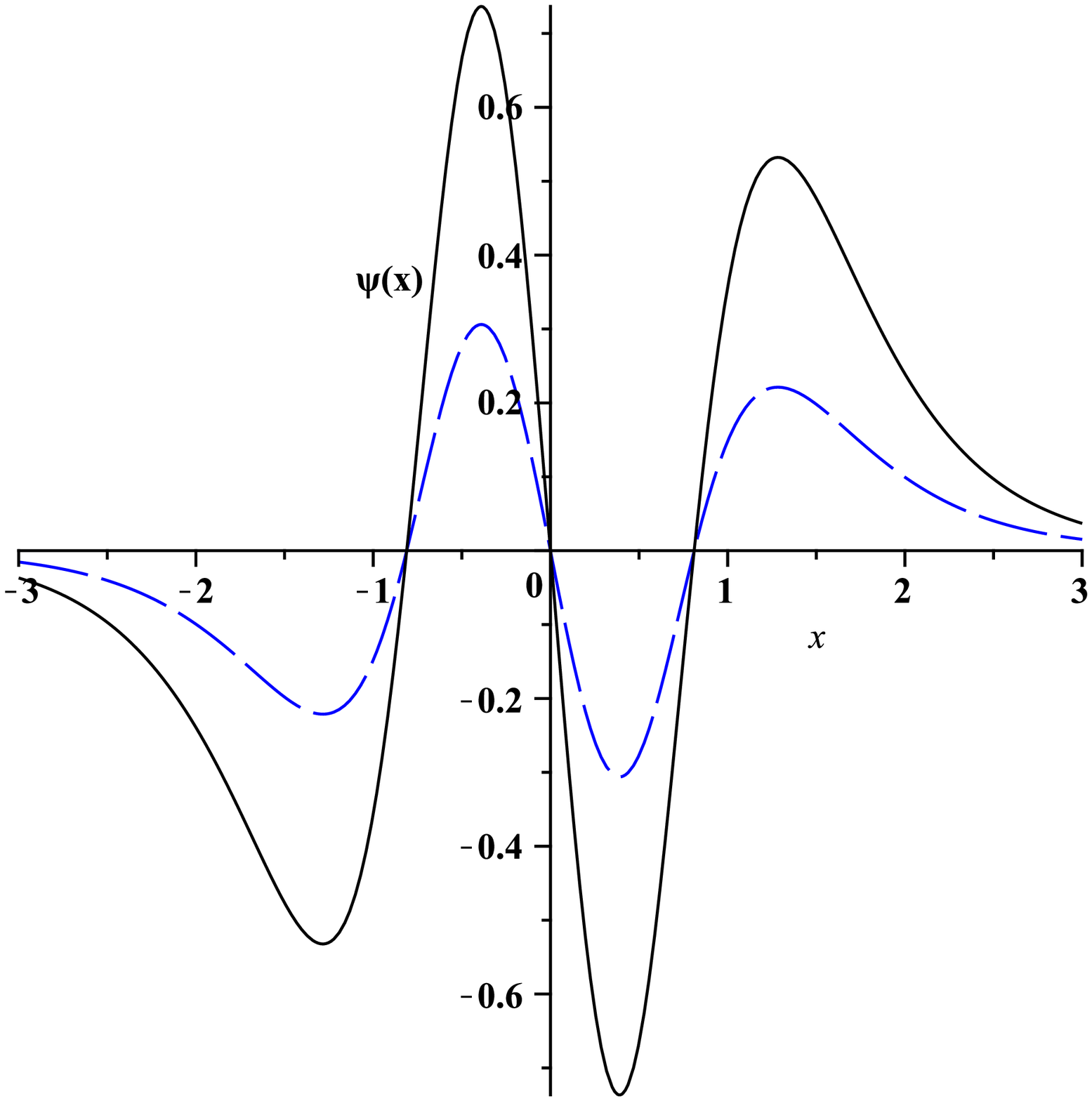}}
{\includegraphics[width=5.cm,height=4.5cm]{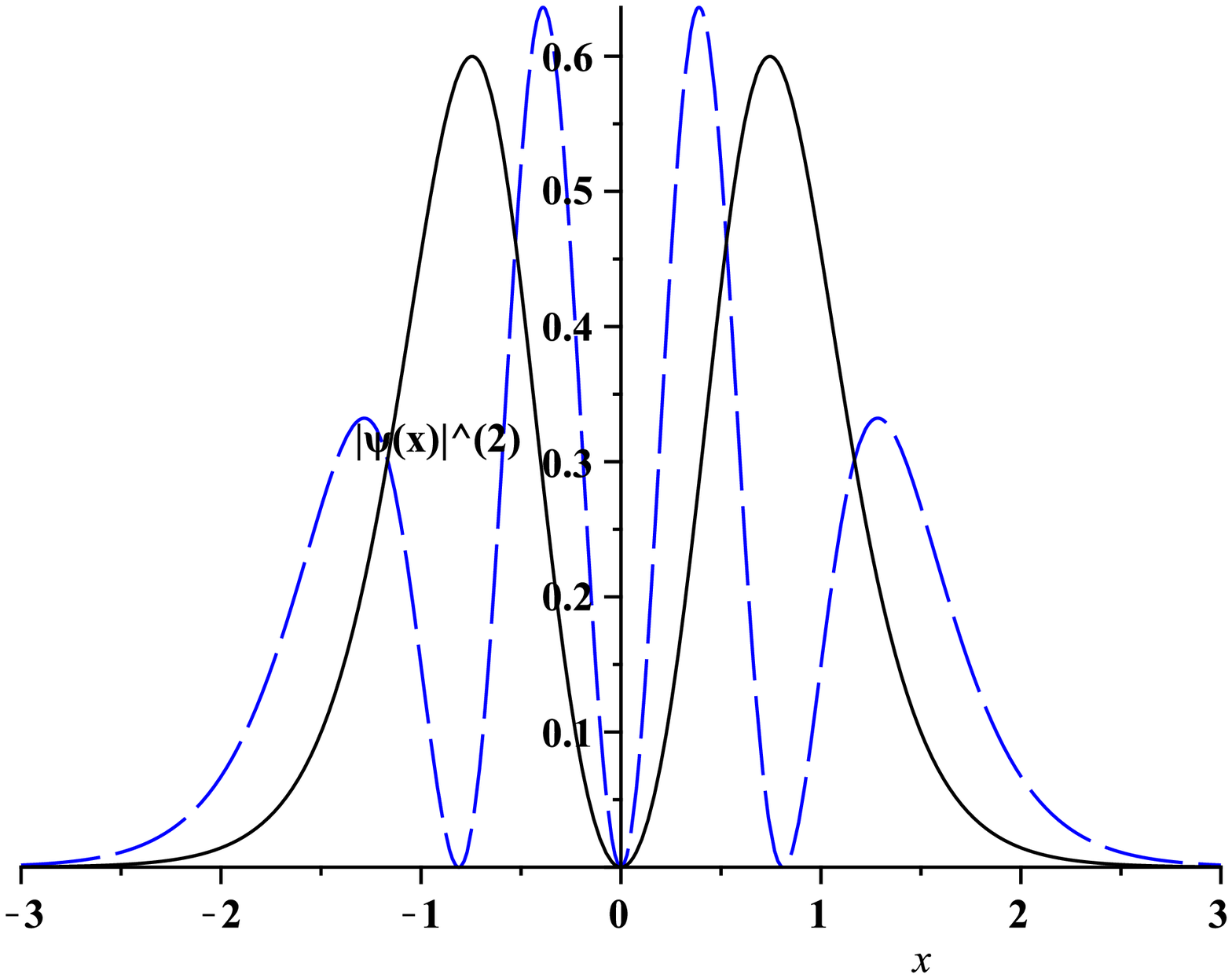}}
\caption{\label{pq_24_V50X} Plot in x-space of normalized real (solid line) and imaginary (dashed) parts of $\psi^{(1)}(x)$, Eq. (\ref{psisim_pq24}), for $\mathcal{E}_0 = -5.210246244135 $ (up left), $\mathcal{E}_2 = 5.7464269736389 $ (up center),  and $\vert\psi^{(1)}(x)\vert^2$ for both energies (up right), and real (solid line) and imaginary (dashed) parts of $\psi^{(2)}(x)$, Eq. (\ref{psiantisim_pq24}), for $\mathcal{E}_1$= -3.793253878015 (down left) and $\mathcal{E}_3$= 13.0906628480 (down center) and $\vert\psi^{(2)}(x)\vert^2$ (down right) for both energies. These correspond to the eigenfunctions $\varphi^{(1)}(z)$ and $\varphi^{(2)}(z)$ for a double-well potential $\mathcal{V}_{(2,4)}(z)$ when $\mathcal{V}_0=50$; see Fig. \ref{pot_Vz_pq_24}.}
\end{figure}


\begin{figure}[h]
\center
{\includegraphics[width=5.cm,height=4.5cm]{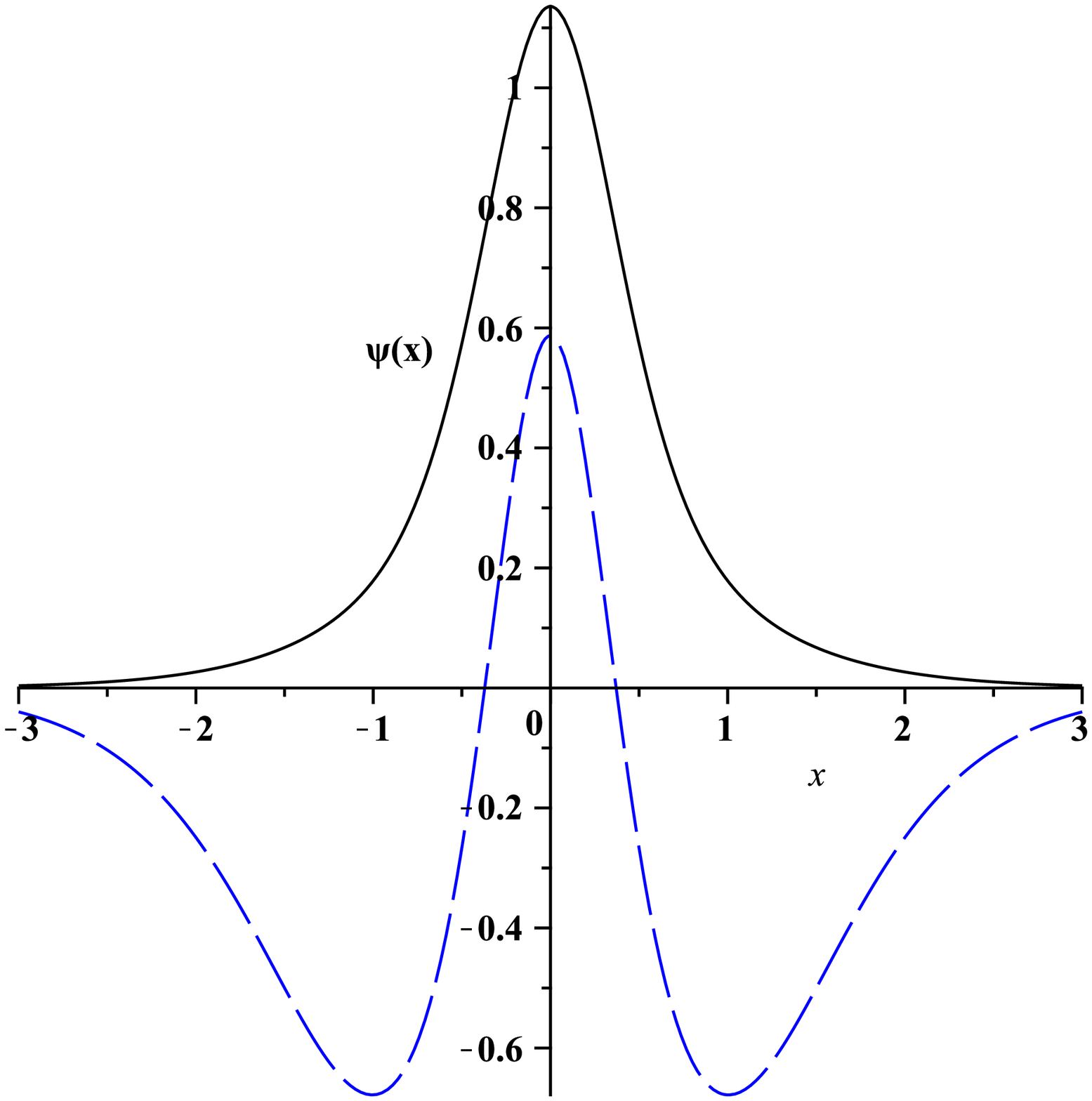}}
{\includegraphics[width=5.cm,height=4.5cm]{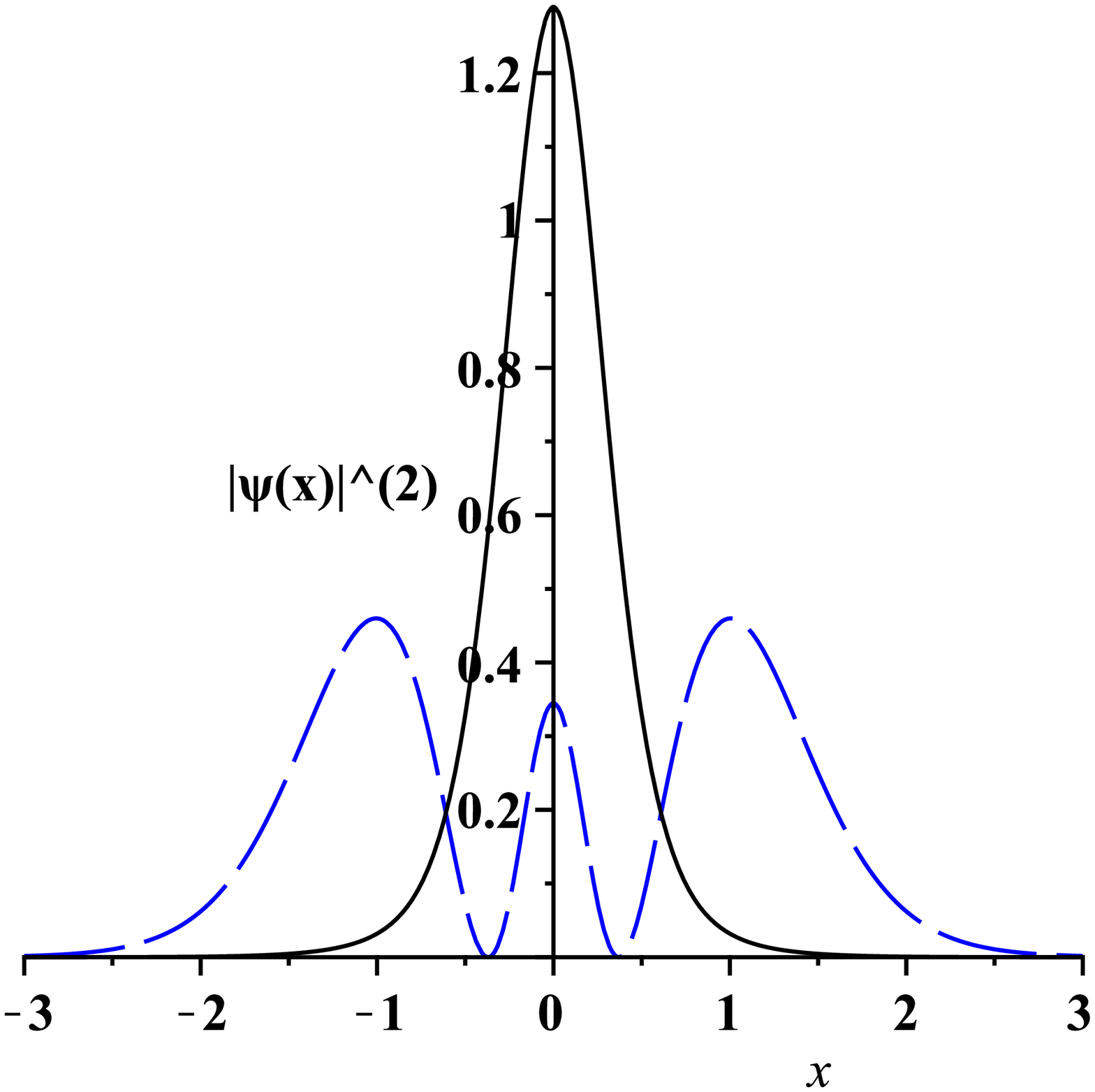}}
\center
{\includegraphics[width=5.cm,height=4.5cm]{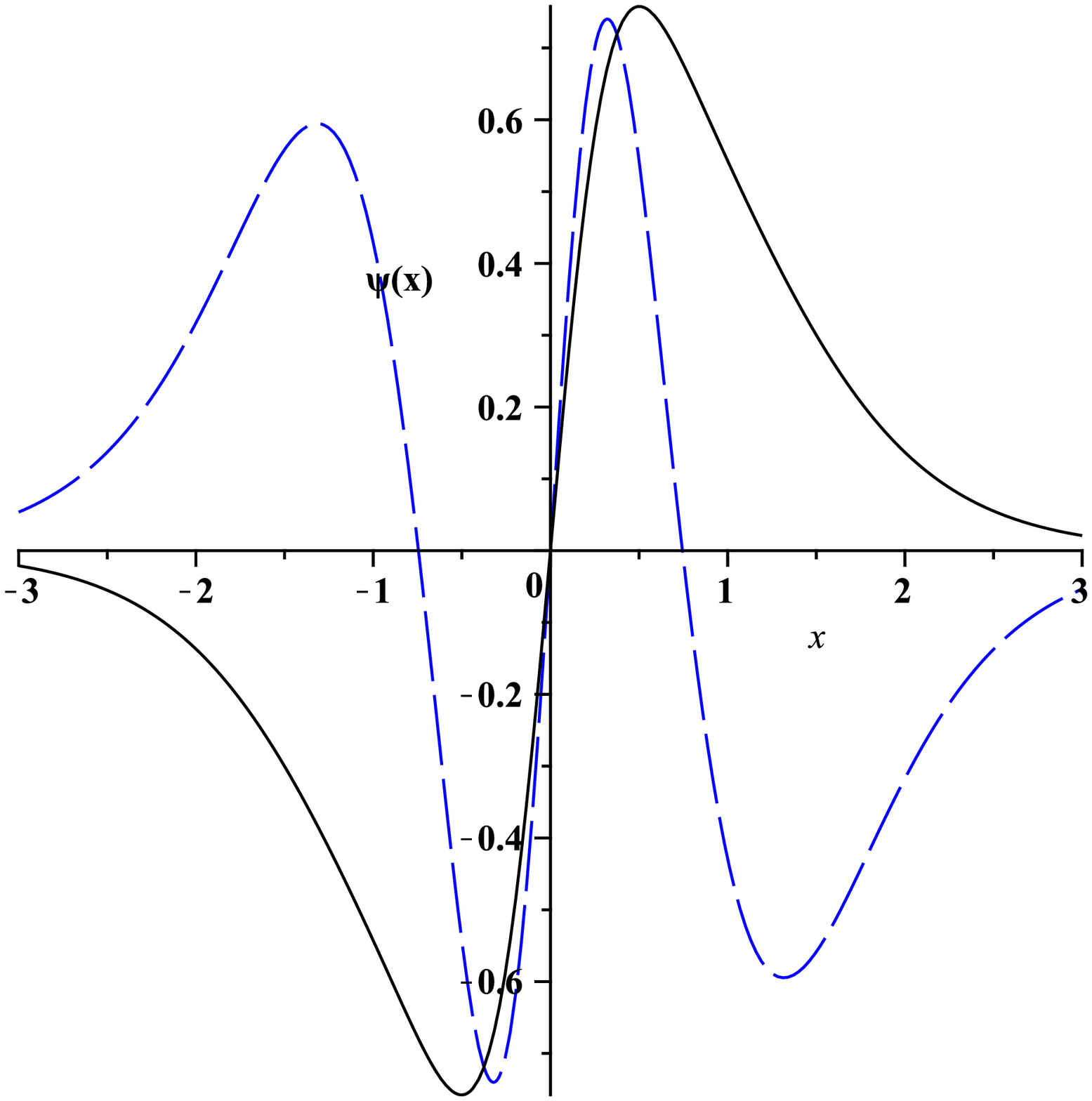}}
{\includegraphics[width=5.cm,height=4.5cm]{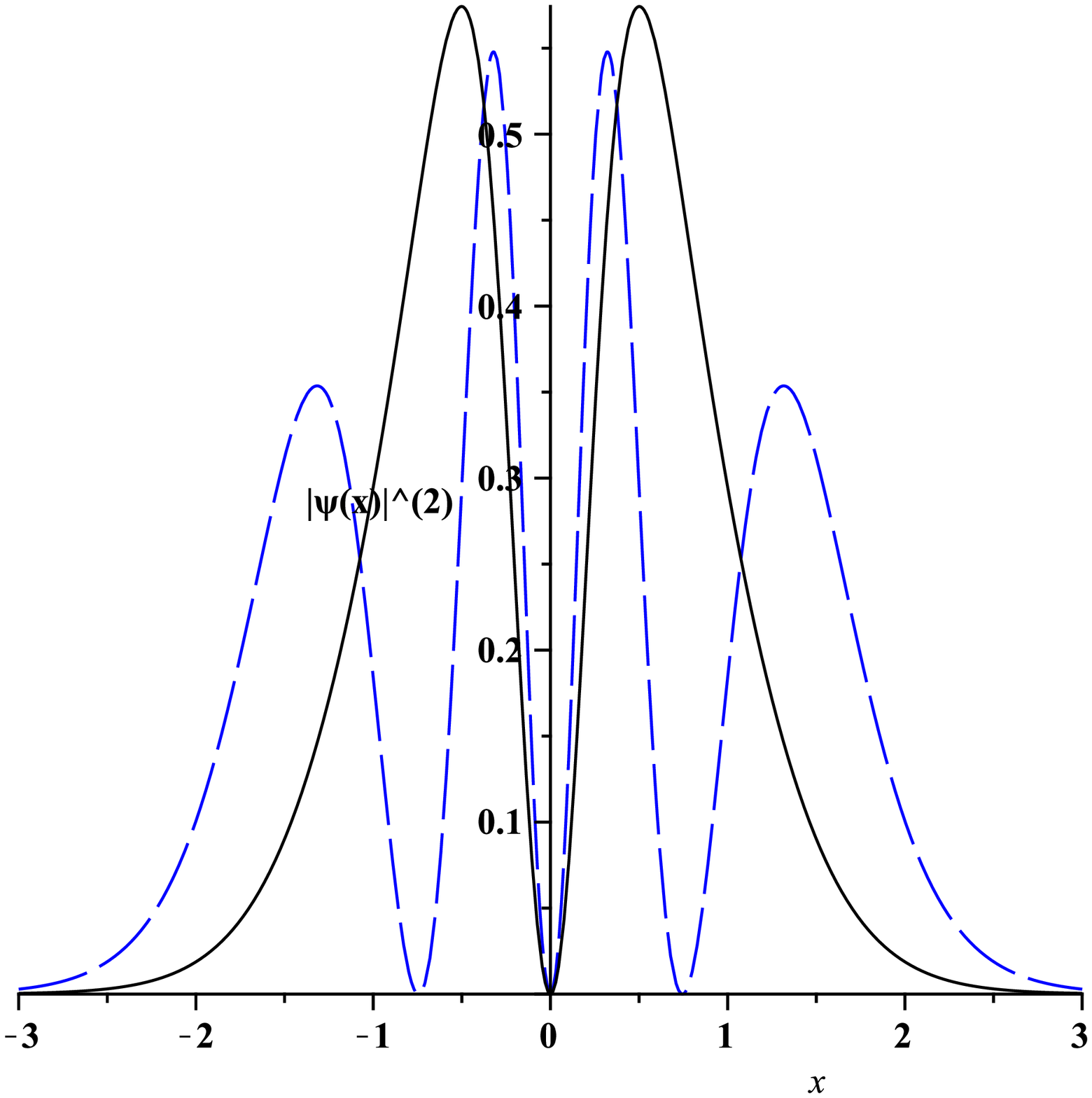}}
\caption{\label{pq_24_V-50X} Plot in x-space of normalized $\psi^{(1)}(x)$ (up left) and $\vert\psi^{(1)}(x)\vert^2$ (up right), Eq. (\ref{psisim_pq24}), for $\mathcal{E}_0$ = 6.46273117442 (solid line), $\mathcal{E}_2$ = 20.09128324147 (dashed), and $\psi^{(2)}(x)$ (down left) and $\vert\psi^{(2)}(x)\vert^2$ (down right), Eq. (\ref{psiantisim_pq24}),  for $\mathcal{E}_1 $= 15.1770345117 (solid line) and $\mathcal{E}_3 $=  26.7184553916 (dashed). These correspond to the eigenfunctions $\varphi^{(1)}(z)$ and $\varphi^{(2)}(z)$ for a triple-well potential $\mathcal{V}_{(2,4)}(z)$ when $\mathcal{V}_0=-50$; see Fig. \ref{pot_Vz_pq_24}.}
\end{figure}

\begin{figure}[h]
\center
{\includegraphics[width=7.cm,height=5.5cm]{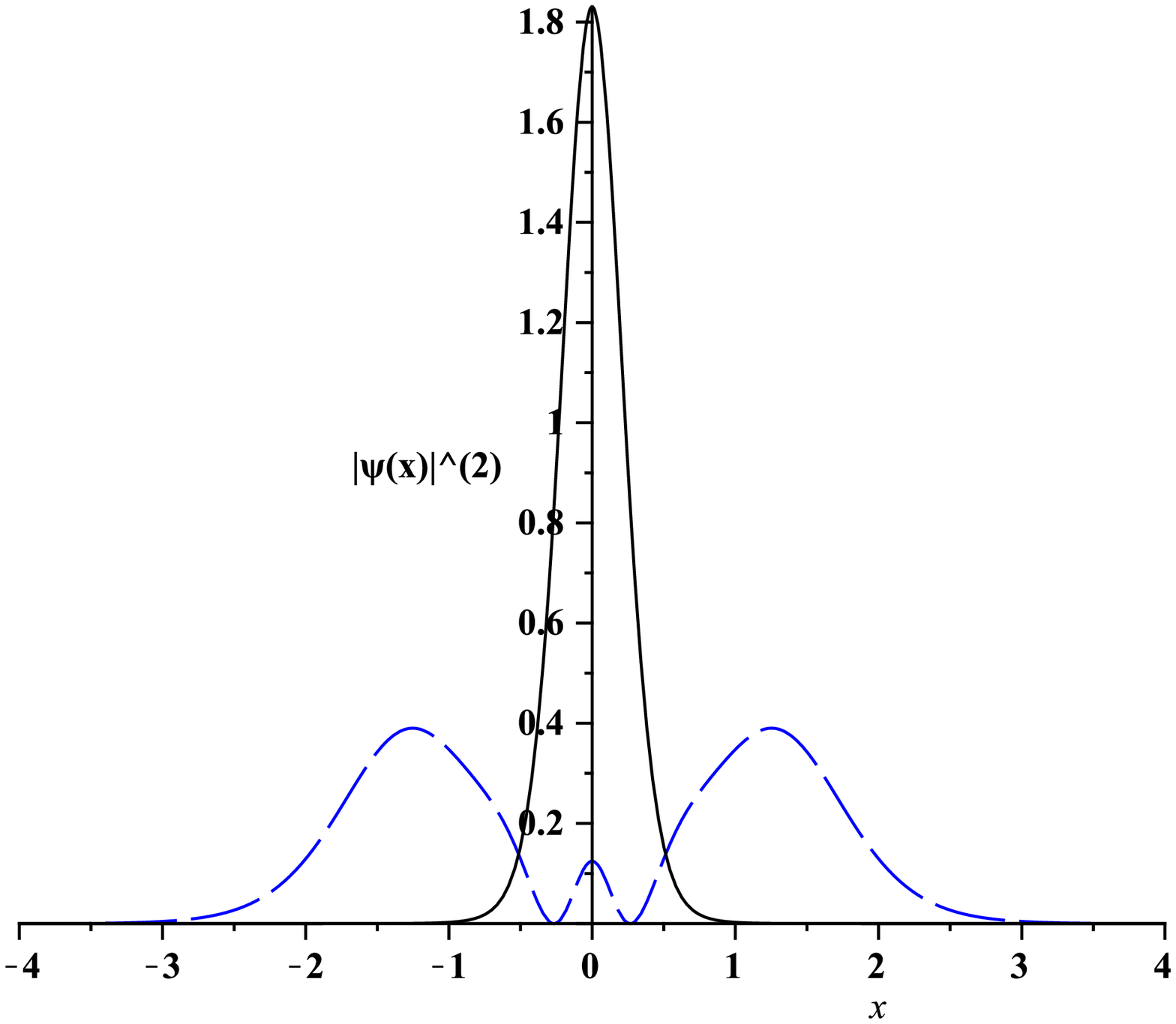}}
{\includegraphics[width=7.cm,height=5.5cm]{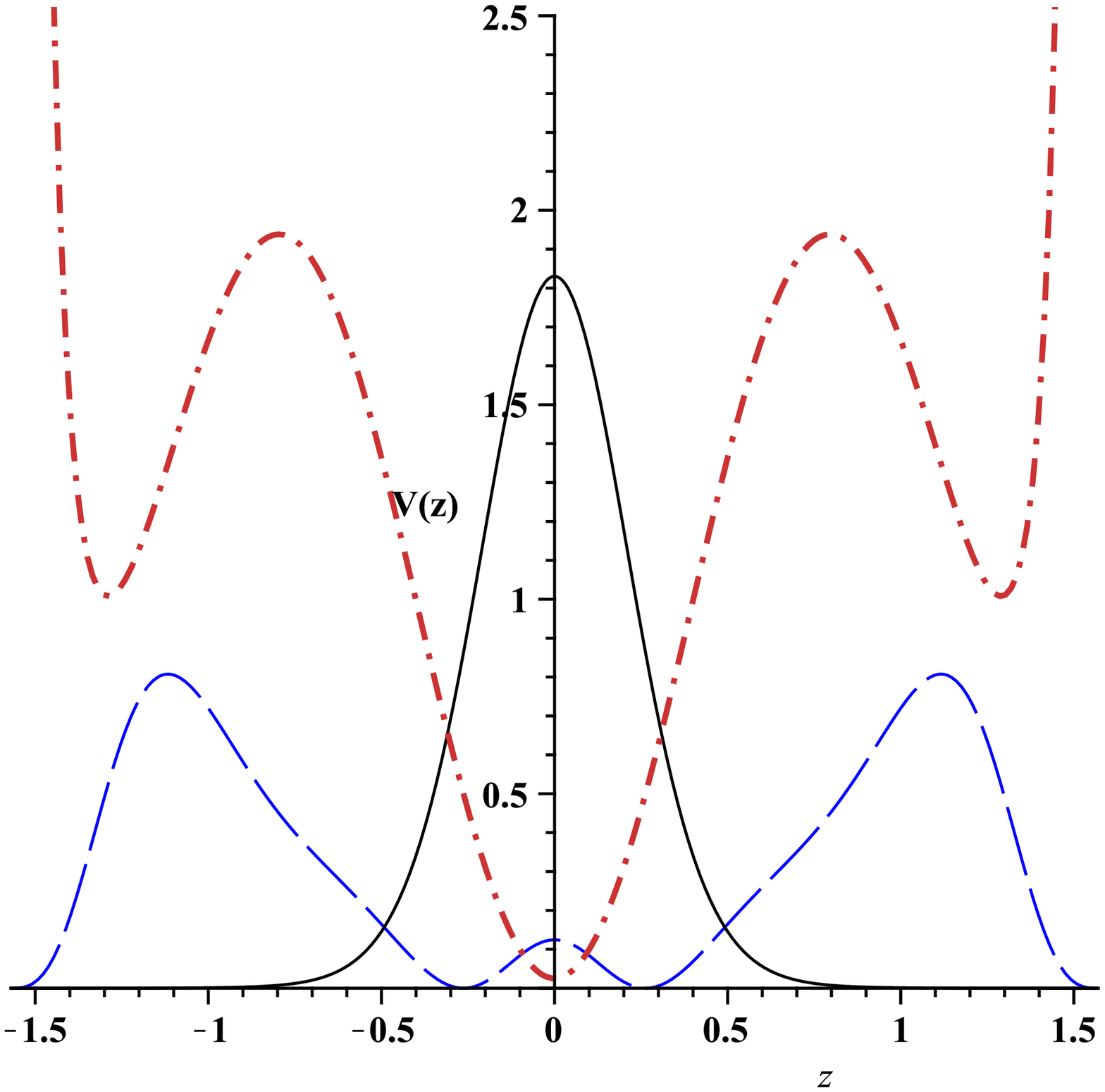}}
\caption{\label{erot_Vz_func_pq_24_Vo-150} (left) Plot of normalized $\vert\psi^{(1)}(x)\vert^2$ in x-space,
Eq. (\ref{psisim_pq24}), corresponding to the eigenfunctions plotted at the right.
(right) Plot of $\mathcal{V}_{{(2,4)}}(z)$ (dot-dashed red line) and $\vert\varphi^{(1)}(z)\vert^2$
for $\mathcal{E}_0 = 11.675703293036$ (solid black) and  $\mathcal{E}_2 = 37.459017784502$ (dashed blue),
when $\mathcal{V}_0=-150$.}
\end{figure}

An analytical survey indicates that the effective z-space potentials vary from double-wells (for $\mathcal{V}_0>3/4$), pass through single-wells (for $-81/4 <\mathcal{V}_0<3/4$), and then to triple-wells (for $-81/4>\mathcal{V}_0$); see Fig. \ref{pot_Vz_pq_24}. The x-space eigenfunctions exhibit the corresponding behavior, similar to the z-shape, as shown by the probability densities displayed in Figs. \ref{pq_24_V50X}, \ref{pq_24_V-50X} and \ref{erot_Vz_func_pq_24_Vo-150}. We could not find a cut for the series in the present case so energy eigenvalues have been computed numerically.

\section{Final Remarks\label{sec:conclusion}}

We have analyzed, in the framework of position-dependent massive quantum particles,
a recently reported family of hyperbolic potentials \cite{downing}.
In the context of the ordinary constant mass Schr\"odinger equation,
the author of \cite{downing} signaled  the role of the confluent Heun functions in the
dynamics of two family members, emphasizing the fact that Heun functions are
rather exotic though increasingly popular in contemporary physics and mathematics.
Indeed, Heun functions have been practically absent in the literature for almost one century
but have been lastly emerging in all frontiers of physics \cite{christiansencunha2011,christiansencunha2012,rumania,bulgaria,cvetic2011,herzog}.
In the present paper, we reconsidered the same class of potentials when the mass is promoted
to a phenomenologically interesting soliton-like distribution. We extended the number of cases analyzed
so that our results are rather comprehensive. Although the differential equations are dramatically modified,
we managed to analytically handle them and found their solutions in several cases.
In fact, the PDM upgrade makes the equations much more difficult and some family members
have no analytical solutions for PDM, e.g. the cases $q=4,6$
which for a constant mass have confluent Heun solutions. Among the soluble members in the PDM framework,
we have identified confluent Heun solutions in most of them. For example,
the cases $(-2,2)$, $(0,2)$ and $(2,2)$
have hypergeometric solutions for a constant mass but become confluent Heun functions for PDM.
We also worked out the \textit{PDM} Poschl-Teller potentials and the \textit{PDM} hyperbolic
double-well for their phenomenological relevance and found both to have confluent Heun solutions.
Finally, we would like to point  the fact that analogue cases belonging to a larger class which admits odd
parameters $p$ and $q$ can also have Heun solutions. For example, in the case
$V_{(1,1)}(x) = -V_0\tanh\!x$, which corresponds to
$\mathcal{V}(z)= \frac{1}{2} +\frac{3}{4}\tan^2(z)-\mathcal{V}_0 \sin(z)$,
solutions read \cite{cunha-christiansen2013}
$\psi(x)=\left({\meio+\meio\tanh{x}}\right)\,Hc(0,1,-1,2\mathcal{V}_0,\meio-\mathcal{V}_0-\mathcal{E};
\meio+\meio \tanh{x}).$




\end{document}